\documentclass[preprint]{aastex}

\usepackage[]{graphicx}

\shorttitle{MMIRS}
\shortauthors{McLeod et al.}

\begin{document}



\title{MMT and Magellan Infrared Spectrograph}


\author{Brian McLeod, Daniel Fabricant, George Nystrom, Ken McCracken, Stephen Amato,
Henry Bergner, Warren Brown, Michael Burke, Igor Chilingarian, Maureen Conroy, Dylan Curley,
Gabor Furesz, John Geary, Edward Hertz, Justin Holwell, Anne Matthews,
Tim Norton, Sang Park, John Roll, \& Joseph Zajac}
\affil{Harvard-Smithsonian Center for Astrophysics, 60 Garden St., Cambridge, MA 02138}
\author{Harland Epps} \affil{UCO/Lick Observatory, University of California, Santa Cruz, CA 95064}
\author{Paul Martini} \affil{Department of Astronomy, The Ohio State University, Columbus, OH 43210}



\begin{abstract}
The MMT and Magellan infrared spectrograph (MMIRS) is a cryogenic multiple slit spectrograph operating in
the wavelength range 0.9-2.4 $\mu$m. MMIRS' refractive optics offer a 6\farcm9 by 6\farcm9 field of view for imaging
with a spatial resolution of 0.2 arcsec per pixel on a HAWAII-2 array.  For spectroscopy, MMIRS can be used with long slits up
to 6\farcm9 long, or with custom slit masks having slitlets distributed over a 4$^{\prime}$ by 6\farcm9
area.  A range of dispersers offer spectral resolutions of 800 to 3000.  MMIRS is designed to be used at the f/5 foci of the MMT or
Magellan Clay 6.5m telescopes. MMIRS was commissioned in 2009 at the MMT and has been in routine operation at the Magellan Clay
Telescope since 2010.  MMIRS is being used for a wide range of scientific investigations from exoplanet atmospheres to Ly$\alpha$ emitters.
\end{abstract}




\section{INTRODUCTION}

\label{sect:intro}  

\begin{figure}
\begin{center}
\begin{tabular}{c}
\includegraphics[height=4in, angle=0]{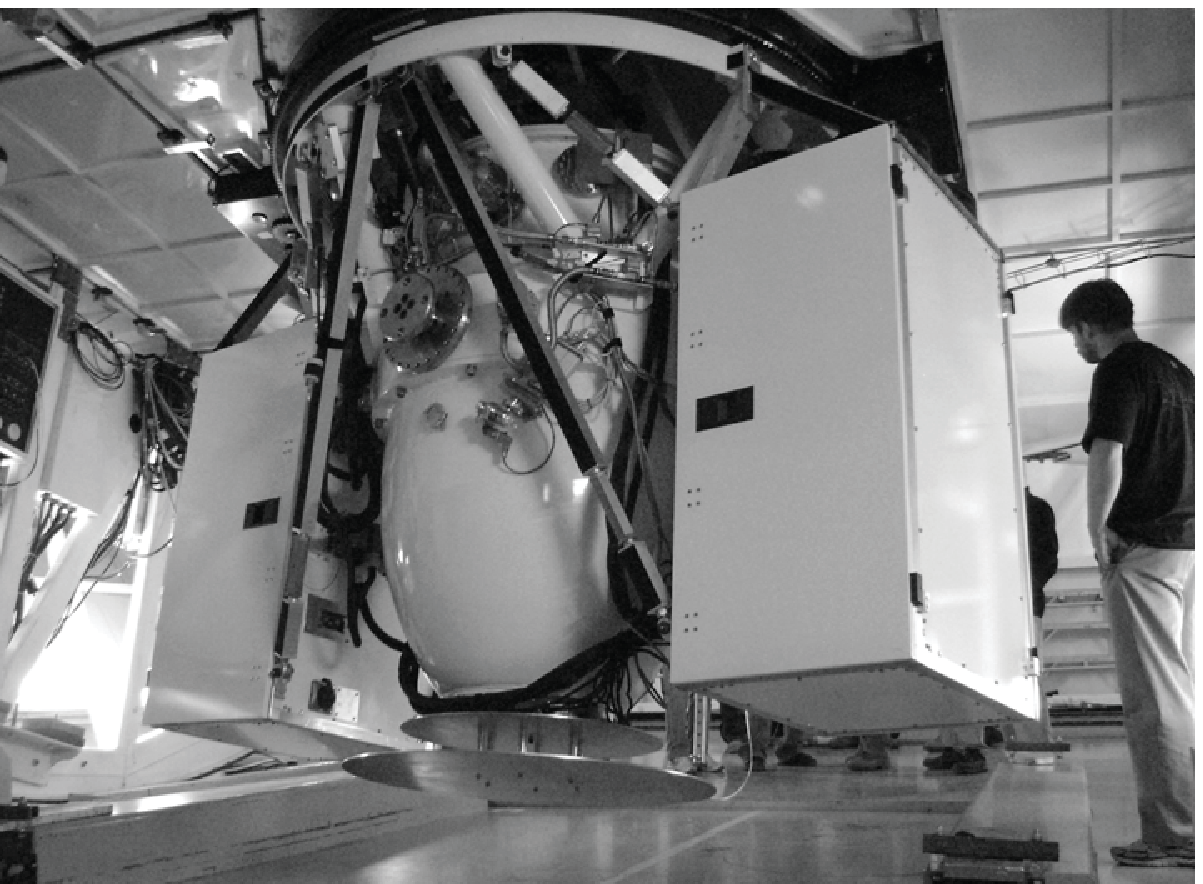}
\end{tabular}
\end{center}
\caption{\label{fig:photo}  MMIRS at the MMT in 2009}
\end{figure}

MMIRS \citep{mcl04} is a cryogenic imaging spectrograph  that can be
used at the f/5 foci of the converted MMT or the Magellan Clay
telescopes.  MMIRS operates in the wavelength range 0.9 to
2.4 $\mu$m and performs spectroscopy with slit masks or long slits.
Commissioned in 2009 at the MMT and Magellan, MMIRS has been in routine operation at
the Magellan Clay telescope since 2010 (see Figure \ref{fig:photo}). The MMIRS design is based on the FLAMINGOS and
FLAMINGOS2 instrument concepts \citep{els98,els03,eik06} developed at the
University of Florida, but MMIRS was independently engineered from the ground
up by our group.

A new generation of infrared multiple object spectrographs is becoming available
at a number of 6.5 to 10 m telescopes.  In addition to MMIRS, these instruments include
MOIRCS at Subaru \citep{suz08}, LUCIFER at LBT \citep{sei10}, MOSFIRE at Keck \citep{kul12},
and soon FLAMINGOS 2 at Gemini \citep{eik06} and EMIR at Gran Telescopio Canarias \citep{gar06}.
MOIRCS was ready for scientific operations in 2006, followed by LUCIFER and MMIRS in 2009, and
MOSFIRE in 2012. The interest in this new generation of spectrographs is based on our conviction that
infrared imaging and spectroscopy are poised to answer a number of high-impact
questions in astronomy, from the formation of the first galaxies and the star
formation history of the universe, to the formation of stars in our own Galaxy.

One of our major goals for MMIRS is developing a physical understanding of high redshift galaxies. For example, galaxy metallicities are estimated from the strong nebular emission lines H$\alpha$ , [NII], [OIII], H$\beta$ , and [OII].  These rest-frame optical emission lines are redshifted in the near infrared for high redshift galaxies, and can be efficiently observed in the redshift ranges:  0.75 $<$ z $<$ 1.05 (H$\alpha$  in J-band),  1.28 $<$ z $<$ 1.74 (H$\alpha$  in H-band, [OIII] in J-band), and  2.0 $<$ z $<$ 2.7  (H$\alpha$  in K-band, [OIII] in H-band, [OII] in J-band). Well documented techniques \citep[e.g.][]{kew02,kew04} allow us to determine gas-phase abundances with [OIII], [OII], and Balmer emission-lines. The metallicity evolution of high redshift galaxies places an important constraint on their star formation history, and relates to mass loss and changing gas fractions in the galaxies.  To a limit of K=20.1 (K$_{AB}$ = 22) the surface density of galaxies in the 1 $<$ z $<$ 2.7 range is roughly 3.5 per sq. arcmin so that the surface density of high redshift galaxies is well matched to MMIRS' multi-slit capability and sensitivity.

MMIRS' broad scientific appeal is demonstrated by the diverse range of observations undertaken in the past 2 years.  In the nearby Universe, MMIRS has been used for studies including transiting exoplanet atmospheres \citep{bea11}, dust chemistry in post-AGB stars \citep{har11}, a planetary nebula in the Sagittarius Dwarf \citep{ots11} and the nature of X-rays sources in the Galactic Center \citep{ser11}.  At cosmological distances, MMIRS has been used to study Ly$\alpha$ emitters \citep{has12}, the physical properties of z$\sim$1.5 galaxies, and high redshift lensed galaxies.

We describe MMIRS' optics in Section 2, its mechanical and thermal design in Sections 3 and 4, and its electrical design in Section 5.
In Section 6 we describe the major challenges that we had to overcome to bring MMIRS into operation.  In Section 7 we describe MMIRS' software
and in Sections 9 and 10 daytime operations and observing protocols.  We conclude in Section 10 with a description of MMIRS' scientific performance.
We use Vega-referenced magnitudes throughout unless explicitly noted.


\section{OPTICAL DESIGN}


\subsection{Overview}

The MMIRS optical train consists of a vacuum-spaced doublet field corrector,
a six-element corrector, and a six-element camera.  The 520mm focal length collimator operates
at f/5.2, producing a 100mm diameter collimated beam.  The 285mm focal length
camera operates at f/2.85 and produces a focal plane scale of 0\farcs201 pixel$^{-1}$
for the 18 $\mu$m detector pixels.  The field of view subtended by the 2048 by 2048 pixel
Hawaii-2 detector is 6\farcm9 by 6\farcm9. The
field corrector produces sharp images across the extended 14$^{\prime}$ diameter
field of view used for guiding.

\begin{figure}
\begin{center}
\begin{tabular}{c}
\includegraphics[trim=1 1 1 1,clip, width=6in]{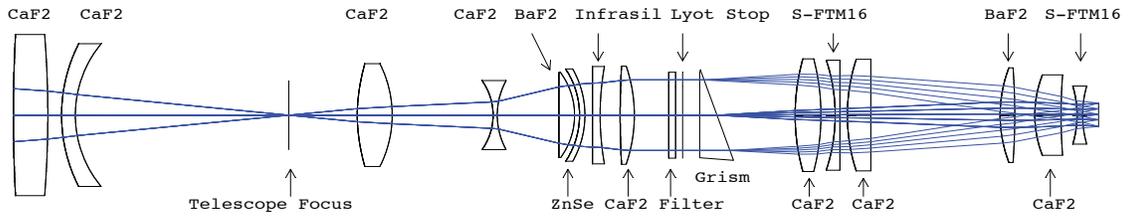}
\end{tabular}
\end{center}
\caption{\label{fig:optics}  MMIRS optics configured in spectroscopic mode.  For imaging, the disperser following the Lyot stop is removed.  The overall distance from the telescope focus to the detector (on the far right) is 1.18 m.  The optics include 8 CaF$_2$, 2 BaF$_2$, 2 S-FTM16, 1 INFRASIL, and 1 ZnSe lenses. All lenses are singlets that are antireflection coated on both sides. The collimator operates at f/5.2 and the camera at f/2.8.}
\end{figure}

The optical layout is shown in Figure~\ref{fig:optics}.
The optical design uses five materials: CaF$_2$, BaF$_2$, ZnSe, Infrasil, and the
Ohara optical glass S-FTM16 \citep{bro04}.  During the optical design process several
constraints were introduced into the optimization operands to control ghost
images. Ghost analysis shows that there are no significant ghost images or
ghost pupils in the MMIRS optics, other than low level reflections from the back of
slit masks and the filters. The optics, including AR coatings, were fabricated
by Janos Technologies.

The final as-built prescription is given in Table~\ref{table:prescrip}.
The monochromatic RMS image diameters (including alignment errors not shown
in Table~\ref{table:prescrip}) averaged over
wavelengths between 0.9 and 2.45 $\mu$m are 11 $\mu$m and 16 $\mu$m at the field
center and corners, respectively.  The worst monochromatic RMS image diameter is
30 $\mu$m at any field angle or wavelength.  The monochromatic 90\% encircled
energy diameters averaged over wavelengths between 0.9 and 2.45 $\mu$m are
18 $\mu$m and 23 $\mu$m at the field center and corners, respectively.
The worst monochromatic 90\% image diameter is
42 $\mu$m at any field angle or wavelength.  The design has a maximum lateral
color of 19 $\mu$m across the full wavelength range at full field, or 12 $\mu$m
across any of the standard photometric bands.
The final scale is 0\farcs2 per 18 $\mu$m pixel.  After our design
was complete improved cryogenic refractive index measurements
by the CHARMS group at Goddard Space Flight Center
\citep{lev05,lev07} became available.  We found that a refocus restored the
original image quality when the CHARMS indices were adopted.
Cold tests of the optics with test masks confirm that we achieve the expected performance.

\subsection{Field corrector}

A doublet CaF$_2$ field corrector lies in front of the f/5 Cassegrain focus of
the converted 6.5m MMT or Clay Magellan Telescope and produces f/5.2 images with
RMS spot diameter less than 0\farcs1 over a 14\arcmin\ field of view.
The design of Cassegrain field correctors is discussed by
e.g. \cite{su90,wil96,epp97}.

The first element of the MMIRS corrector also serves as the vacuum
window of the cryostat.  The 50\,mm thickness of this element limits its internal stress
to 125 PSI under atmospheric pressure.
The second corrector element is inside the cryostat but cools only slightly by radiation
to the cold mechanisms below.  The second element serves as a radiation shield for
the first element, keeping the first element close to the ambient temperature and preventing
condensation from forming.

\subsection{Collimator and Lyot stop}

The 520mm focal length collimator has 6
elements, 3 made of CaF$_2$, and one each of BaF$_2$, ZnSe, and Infrasil.
At f/5.2, the collimator produces a 100mm
diameter collimated beam with a pupil ~65\,mm from the back vertex of the
final collimator lens.  The pupil blur over the full range of field
angles and wavelengths is $\pm$2.5 mm.  The physical Lyot stop is placed
just beyond the optimal pupil position due to physical constraints.

The leading surface of the first collimator lens is aspheric with
4th, 6th and 8th order terms and a maximum aspheric departure of 0.19 mm.
This surface was diamond-turned, post-polished, and tested with subaperture
stitching interferometry.

The Lyot stop was laser cut from 0.2mm thick anodized aluminum foil by
PhotoMachining of Pelham, NH.  The 33mm diameter central obscuration
in the stop is supported by 0.75mm wide spider arms.  These were made
as thin as possible to minimize the amount of blocked light.  Because the
Lyot stop does not rotate, the spider will not in general line up
with the telescope spider as the telescope tracks.

\subsection{Filters and dispersers}

The 125\,mm diameter, 10\,mm thick filters are located just before the
Lyot stop.  We currently have seven broadband filters: Y, J, H, K$_s$, K$_{spec}$, zJ, and HK.
The first four are the standard imaging and spectroscopic filters, augmented by the
latter three for spectroscopy.  Table \ref{table:filters} summarizes the filter
characteristics and vendors.

The disperser wheel follows the Lyot stop.  We currently have three surface-relief
grisms: J, H, and HK, all obtained from the Richardson Grating Laboratory.  H and K volume phase holographic
gratings (produced by Kaiser Electro-Optics) with beam straightening prisms (produced by ISP Optics)
will be installed in 2013.  Table \ref{table:dispersers} lists the disperser details.
The available spectroscopic observing modes are summarized in Table \ref{table:modes}.

\subsection{Camera}

The 280mm focal length camera was based on an
earlier design for FLAMINGOS-2 \citep{epp03}.
The f/2.8 camera contains 6 elements, 3 made of CaF$_2$, 2 of S-FTM16, and one of BaF$_2$.
The first infrared camera design using S-FTM16 known to the authors was
a 5 element system developed by Stephen Shectman in 1999 (private communication).

\subsection{Guider and wavefront sensing optics}
\label{sect:guider}

An annular tent mirror directs light outside the science field (to a maximum radius of 7\arcmin) to two
opposing guider/wavefront sensor assemblies.  The light reflected from the
tent mirror passes through two opposing vacuum windows, and the guiders are located
outside the dewar.  The tent mirror was diamond-turned from a single piece of
stress-relieved aluminum 6061 and post-polished.

The guider optics are designed to work from 6000 {\rm \AA} to
9000 {\rm \AA} because MMIRS is primarily a bright-time
instrument.  Each guider assembly is mounted on a 3-axis stage (x, y, and focus).
The focal plane is relayed 1:1 onto the guide cameras using a symmetric pair of
custom achromatic lenses.  The field of view of each guide camera is
1\farcm3$\times$1\farcm3.

In normal operation, one camera is used for guiding, and the other
operates as a wavefront sensor, feeding images to the observatory
active optics system \citep{sch03,pic04}.  To use the guider as a
Shack-Hartmann wavefront sensor, a stage containing two fold mirrors
and the wavefront sensor optics is moved into the beam. The first fold
directs the light through a 4$^{\prime\prime}$ field stop that reduces
sky background. The beam is then folded through a second collimator
that forms an 8mm diameter pupil image on an Adaptive Optics
Associates lenslet array with a 0.6mm pitch and 40\,mm focal length. A
final fold mirror directs the image to the CCD camera.  No refocus is
required when switching between guiding and wavefront sensing.  A
20$\mu$m pinhole illuminated by an LED is used to calibrate the
Shack-Hartman spot positions. The pinhole assembly is attached to the
guider baffle inside the cryostat and the guider assembly is
translated in front of the illuminated pinhole for calibration.  Stars
brighter than R=16 allow guiding with 1 s cadence, and stars brighter
than R=14.5 allow wavefront sensing with 30 s integrations.

\subsection{Calibration optics}
\label{sect:calsys}
The calibration system is housed in a compact system that mounts directly above the main instrument.
The light source is a LabSphere 250\,mm integrating sphere with a Spectralon coating.  The 50 mm
exit port of the sphere is conjugated to the Lyot stop using a Fresnel Technologies POLY-IR5
IR-transmissive Fresnel lens.  The system illuminates
the full 7$^{\prime}\times$7$^{\prime}$ field with an f/5 beam. Two 25\,mm ports on the sphere
contain continuum lamps and an Ar Penray line lamp.  During calibration a polished aluminum
fold mirror translates into the beam to send light to the instrument below.


\section{Mechanical and Thermal Design}


\subsection{Mechanical and thermal overview}

\begin{figure}
\begin{center}
\begin{tabular}{c}
\includegraphics[width=6in]{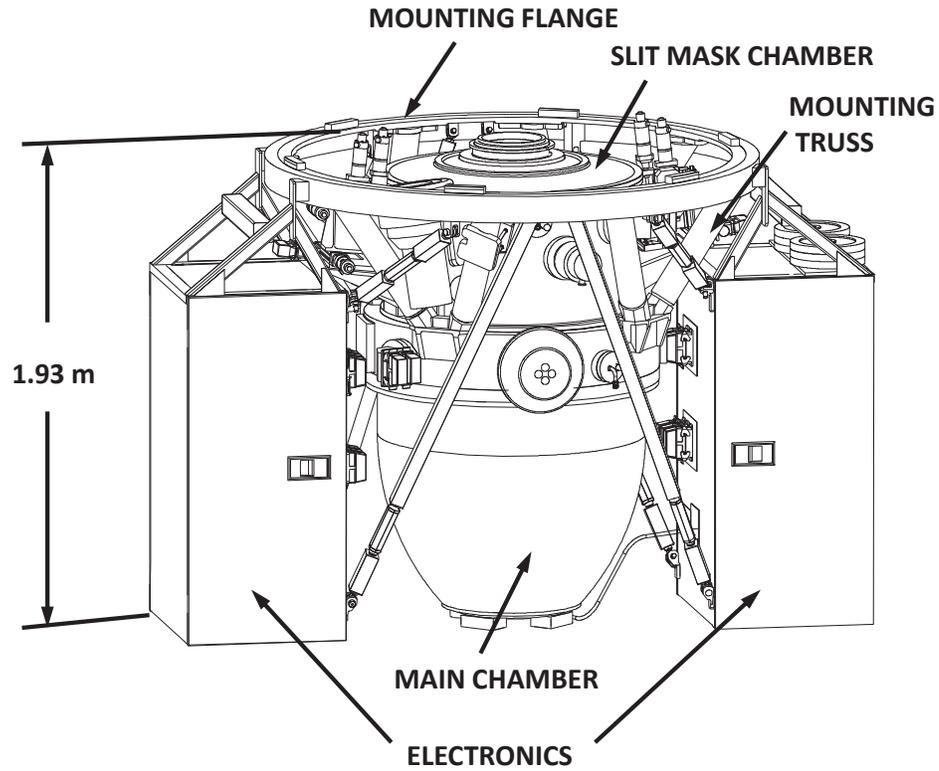}\\
\end{tabular}
\end{center}
\caption{Exterior view of MMIRS.  The overall height of the instrument is 1.93 m, and
the distance from the top of the corrector to the detector is 1.6 m. The two electronics racks are
mounted directly to the mounting flange.  \label{fig:mmirs_layout}}
\end{figure}

\begin{figure}
\begin{center}
\begin{tabular}{c}
\includegraphics[width=6in]{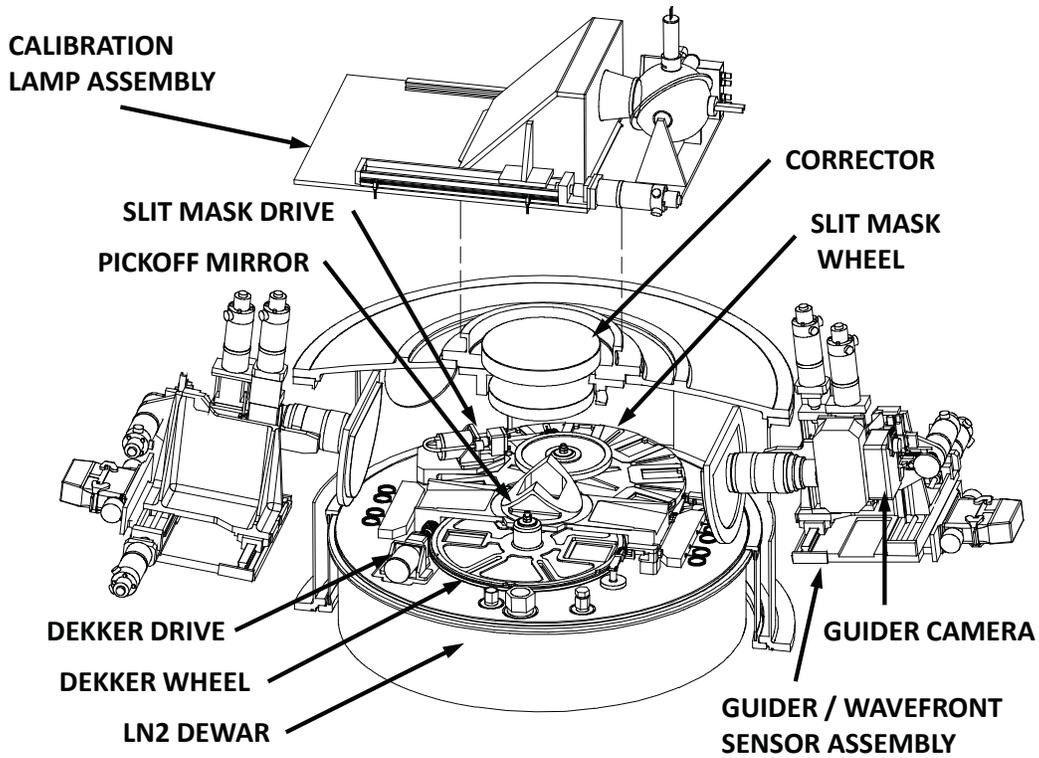}\\
\end{tabular}
\end{center}
\caption{Layout of the slit chamber.  The calibration lamp assembly is shown in the calibration position; during observation
the fold mirror assembly is retracted to the left.  Two identical guiders view through windows on opposite sides of the
instrument.  This chamber can be thermally cycled during the daytime to allow mask exchange.
  \label{fig:slit_chamber}}
\end{figure}

Figure~\ref{fig:mmirs_layout} is an external view of MMIRS.  The total instrument mass
is $\sim$2 metric tons.  The slit mask chamber shown in Figure~\ref{fig:slit_chamber} is
a vacuum vessel at the top of the instrument sealed at the top
by the first corrector lens and at the bottom by a gate valve.
The slit mask chamber and all of its contents can be disconnected from the bulkhead
that forms the top of the main instrument chamber.
The slit mask chamber contains the corrector optics, the Dekker wheel for aperture selection,
the slit mask wheel, a dewar assembly, and the guider pick off mirror.
The gate valve, located between the slit mask wheel and the first collimator lens,
isolates the optics and detector during slit mask exchange.
The gate valve is mounted to the top bulkhead of the main instrument chamber.

The toroidal slit mask LN$_2$ reservoir is supported by an insulating G-10
ring from the bottom flange of the slit mask chamber.  The top of the slit mask
LN$_2$ reservoir serves as the mounting plate for the slit and Dekker wheels
and the guider pick off mirrors.  To the sides are two windows that pass the guider light
outside the cryostat.  The cold baffle that blocks thermal emission from the
warm gate valve is mounted to the central wall of the slit mask chamber LN$_2$ reservoir.

\begin{figure}
\begin{center}
\begin{tabular}{c}
\includegraphics[width=6in]{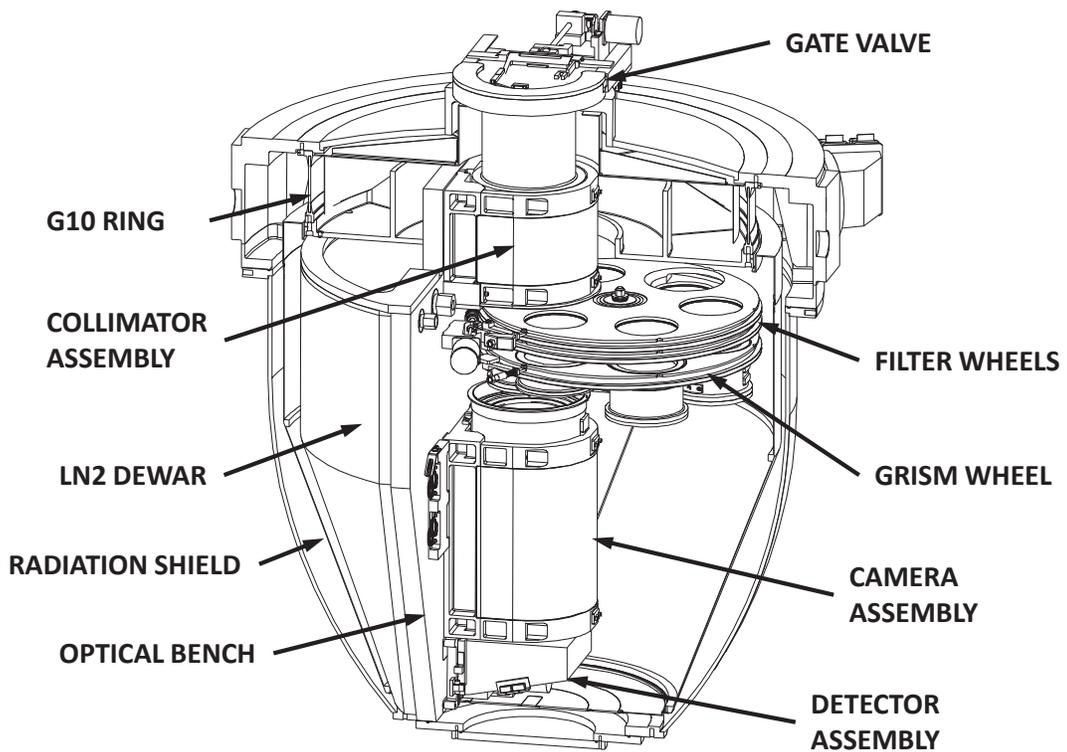}\\
\end{tabular}
\end{center}
\caption{Layout of the main chamber. The main chamber contains the collimator, camera, dispersers and filter.
The main chamber is isolated from the slit chamber by a gate valve when masks are exchanged.\label{fig:main_chamber}}
\end{figure}

The main chamber (Figure~\ref{fig:main_chamber}), sealed at the top by the gate valve,
contains the collimator optics, Lyot stop, filter and disperser wheels, camera optics,
and detector assembly. A second G-10 ring mounted on the inside of the main chamber bulkhead
supports the D-shaped LN$_2$ reservoir in the camera section.
The face of this reservoir is the main optical bench supporting the
collimator and camera optics, disperser and filter wheels, and the
detector assembly.  All major aluminum cryogenic components in MMIRS were stress relieved
by cooling from room temperature to 77K at several steps during the
machining process.

MMIRS contains ten mechanisms: a calibration fold mirror slide, two guider assemblies, a Dekker wheel,
a slit mask wheel, a gate valve (and its baffle), two filter wheels, a disperser wheel, and
a detector focus assembly.  The basic characteristics of these mechanisms are summarized in
Table~\ref{table:mechanisms}.  All of the internal mechanisms are driven by 200 step
revolution$^{-1}$ Phytron VSS-52 vacuum rated stepper motors with 0.35 N-m of
holding torque.  The external guider mechanisms are driven by normal duty Phytron
ZSS-52 stepper motors with 0.45 N-m of torque.

Effective thermal design of MMIRS requires balancing competing factors: (1) to minimize dark current the detector should be as cold as possible, (2) to avoid damage the optics and detector cannot be cooled or warmed too rapidly (3) the slit masks need to be warmed and cooled as rapidly as possible to allow convenient daytime exchange and (4) the system should be as simple as possible. Our approach was to build MMIRS around two liquid nitrogen dewars which are coupled as tightly as possible to the spectrograph components. We control the detector and optics warmup and cool down rates by regulating heater power or the flow of LN$_2$.  To validate this approach we constructed a detailed thermal finite element model of MMIRS to verify cool down and LN$_2$ consumption rates.

Radiation shields surround the cold spectrograph components.  The radiation shields are 1.26\,mm thick
aluminum 1100 sheet metal supported with G-10 standoffs for thermal isolation
from the warm and cold surfaces. A single layer of aluminized Mylar covers
both sides of the G-10 rings, the radiation shields and much of the LN$_2$ tank surfaces.
Electrical connections pass though connectors mounted on the thermal shield flange.
The flange connectors allow easy shield removal and provide a light-tight shield penetration.
Multiple sensors in each chamber provide temperature information.

\subsection{Guider assembly}

The two guider translation axes and the wavefront sensor translation axis are
linear stages mounted on THK rails and driven by ballscrews.  Stepper motors
drive the ballscrews through 6.25:1 planetary gearheads, providing a 4.8 $\mu$m
stepsize.  The guider focus mechanism uses similar rails, encoders, and gearboxes,
but is cam driven over a 2\,mm range with 1.8 $\mu$m resolution.

The guider CCD cameras are based on Steward Observatory and Carnegie Observatory designs.
The guider electronics are adapted from a Carnegie Observatory design, repackaged
to allow the CCD head to be separated from the readout electronics.
The readout electronics use five printed circuit boards (power, clocks, signal processing,
timing generator, and amplifiers), interfaced to a small backplane.  We house this system in
an aluminum box measuring 160$\times$95$\times$85 mm.  Input of control signals and output of video
data is via a fiber optic link to an interface card on the PCI bus of the host computer.
Internal voltages derive from a single 48V DC power input, and total power dissipation for
the system is approximately 12 watts (excluding thermoelectric cooling).

We use an E2V CCD47-20 CCD, a thinned backside-illuminated frame-transfer device with 1024$\times$1024
active pixels (13$\mu$m square).  The CCD is housed in a vacuum-tight aluminum box measuring
approximately 120$\times$85$\times$45\,mm (not including the small vacuum valve, electrical connector,
and heat extractor).  Internal cooling of the CCD to -20C or lower is provided by a 3-stage
Melcor thermoelectric cooler, with the heat being transferred through the back of the housing
to a small fan-cooled heat sink.  Maximum power for the TE cooler is 12 W.

The remote CCD head connects to the readout electronics box via a multi-coax cable.  To better
buffer the CCD output signals for remote operation, small preamplifiers have been added inside
the head for the two video channels.  Tests have shown no performance degradation with cable
lengths of up to six feet.

\subsection{Slit mask and Dekker wheels}

\begin{figure}
\begin{center}
\begin{tabular}{c}
\includegraphics[width=6in]{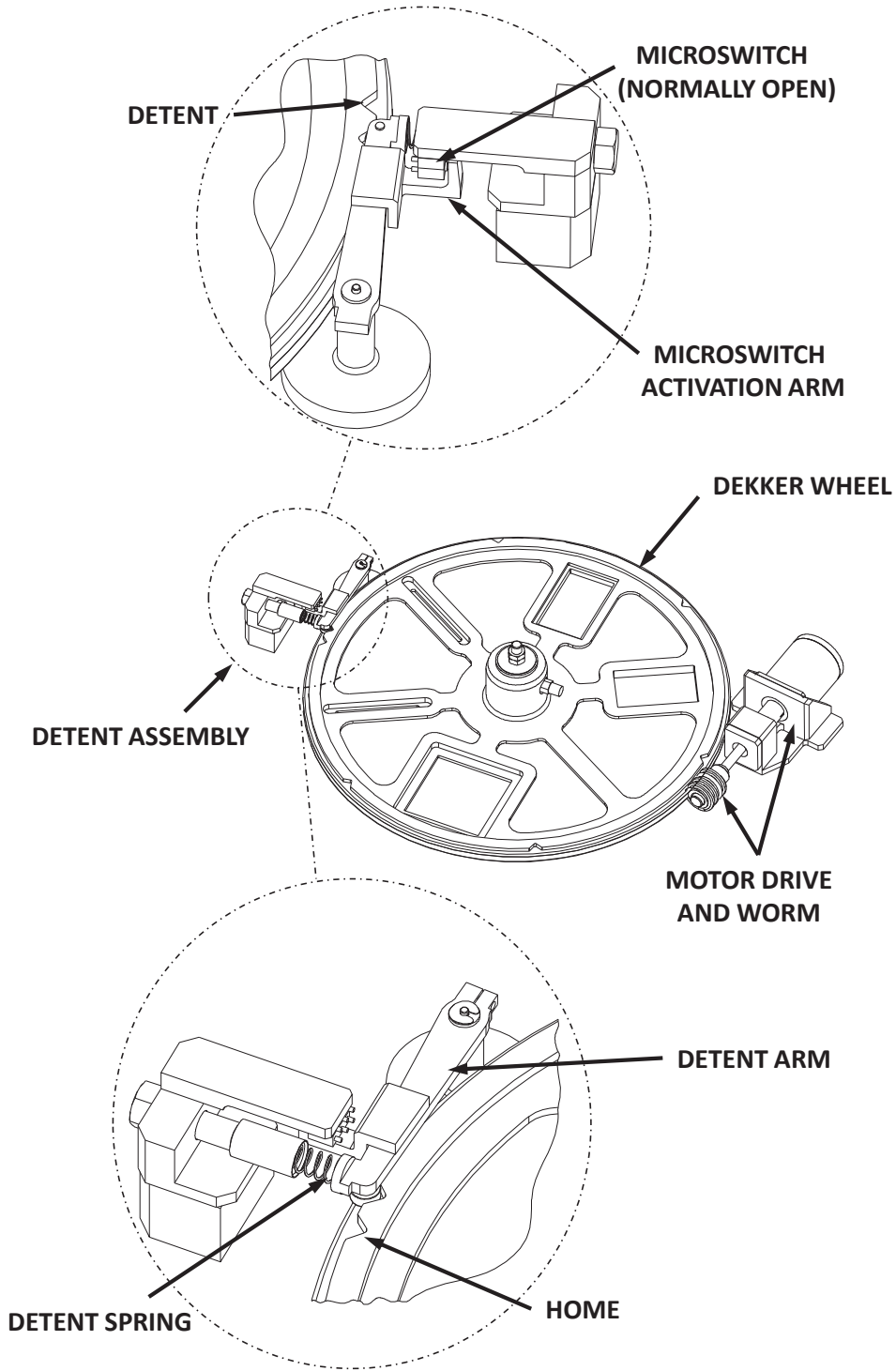}\\
\end{tabular}
\end{center}
\caption{Drive and detent mechanisms for the Dekker wheel.  A stepper motor drives the wheel through a worm gear.
An arm on the detent assembly depresses a microswitch to sense detent positions.  The slit mask,
filter, and disperser wheels operate in the same fashion. \label{fig:wheel}}
\end{figure}

The Dekker wheel is used to limit the field of view for spectroscopy (see Figure~\ref{fig:wheel}). The wheel carries
five apertures: one for long slits, one for short slits, one for slit masks,
a blocking mask, and an open aperture for imaging.
A stepper motor drives a Vespel (polyimide) worm which engages a gear machined into the outer edge of the wheel.  Wheel positions
are selected with mechanical detents (also at the wheel edge) that are sensed with a microswitch.
There is enough backlash in the drive gear to allow the detent to seat fully and control
the wheel positions.  An extra detent position located near the open aperture detent provides a
unique sequence of detents for home sensing at power up. A sapphire ball thrust bearing creates a thermally
conductive path from the wheel to the liquid nitrogen dewar.  The Dekker wheel and slit mask
wheel are primarily cooled through a Vel-therm conductive path at the rotary shaft as discussed
in Section 6.1.

The slit wheel mechanism is similar to the Dekker wheel but it contains
a square imaging aperture, 7 long slits, varying in size from 0\farcs2
(1 pixel) to 2\farcs4 (12 pixels), and 9 multi-slit masks, each
4\arcmin$\times$7\arcmin.  The slit wheel is driven through a gear having
40\% of the diameter of the wheel to avoid interfering with the exchange of
the slit masks.  A port on the vacuum vessel allows radial insertion/withdrawal of the slit
mask holders for mask exchange.

\subsection{Slit mask dewar}

The liquid nitrogen dewar has fill and vent lines arranged to limit LN$_2$ loss
at extreme telescope orientations.
The full slit mask dewar volume is 55 liters, but the fill is limited to 45 liters by the
fill and vent port locations, yielding a hold time of 40 hrs.

The stiffened front surface of the slit mask dewar is the mounting surface for the Dekker wheel,
the slit mask wheel and the guider pick off mirror.  A G-10 tube assembly structurally connects the dewar to
the vacuum vessel. The G-10 tube is epoxied into aluminum mounting rings using
Stycast 2850-FT epoxy with a 0.25mm bond thickness. The G-10 tube minimizes the conductive heat
loss from the dewar and provides compliance to accommodate the thermal contraction of the
the dewar.

\subsection{Gate valve}

The gate valve is a slightly customized version of a VAT Inc.~catalog item.  The
40mm thick gate valve is thin enough to fit in the 100mm gap between the slit
and the first collimator lens. The VAT valve components are
combined with a custom stepper motor driven ballscrew assembly.  The stepper motor
drives the ballscrew through a worm gear. Microswitches sense the closed and open positions.

\subsection{Disperser and filter wheels}

Both six-position filter wheels carry five filters with one open aperture.
Their design is very similar to the Dekker wheel described previously.
The disperser wheel is a six-position wheel intended to carry five dispersers with
one permanent open aperture.
The drive and detent scheme for the filter and disperser wheels is very similar to the Dekker
wheel described previously.  No Vel-therm supplemental conductive path was required in the
disperser and filter wheels because they are cooled quite slowly to avoid stressing the
optics. The disperser wheel is balanced with counterweights to minimize the required drive torque.

\subsection{Focus stage}

\begin{figure}
\begin{center}
\begin{tabular}{c}
\includegraphics[width=6in,angle=0]{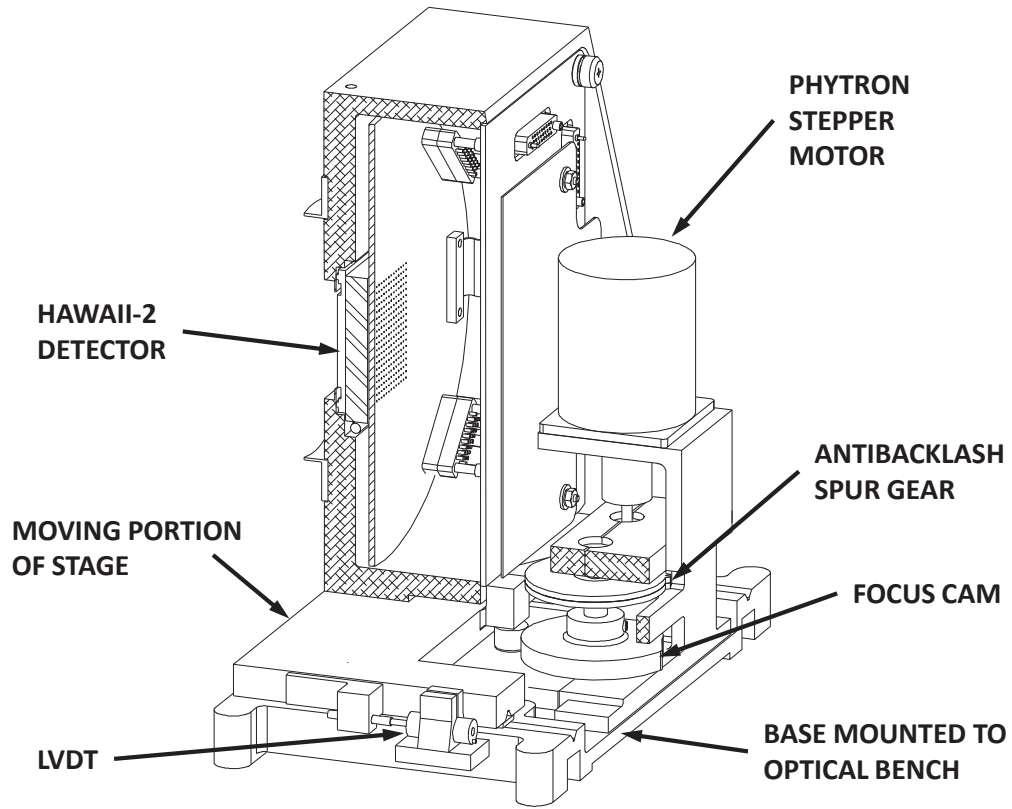}\\
\end{tabular}
\end{center}
\caption{\label{fig:detector} The detector assembly with focus stage.  The HAWAII-2 array detects light
incident from the left. For focus, a stepper motor positions a linear stage
by rotating a cam through reduction gears.}
\end{figure}

The detector focus mechanism is a linear stage with a cam drive shown
in Figure~\ref{fig:detector}.  The cam is stepper motor driven through
a gear reduction.  The stage slides on sapphire balls in machined
raceways that are spring loaded together. The stage is spring loaded
against a cam drive that provides a 5-micron motion per motor step
with a range of $\pm$2.5mm. An LVDT provides position readout with
1$\mu$m resolution.  The stage is cooled primarily via 0.25\,mm thick
copper foil straps that connect the detector housing to the fixed
portion of the stage.  In practice the focus stage is rarely moved as
the spectrograph is quite stable over extended time periods.

\subsection{Main dewar and optical bench}

The optical bench/dewar assembly is used to cool and support the components in
the main chamber. The assembly is composed of four main structural elements.
(1) The front bulkhead is a deep section forging that provides the mounting
interface to the telescope truss. The gate valve separating the slit mask chamber from
the main chamber is mounted on this bulkhead. LN$_2$ and electrical feedthroughs
penetrate the bulkhead.  (2) A G-10 ring assembly with aluminum end fittings
supports and thermally isolates the cold components.  It is similar to the
G-10 ring in the slit mask chamber.  (3) A deep section support ring connects the G-10 ring
to the optical bench/LN$_2$ dewar assembly.
(4) The optical bench/LN$_2$ dewar consists of a 50\,mm thick light weighted top plate welded to a curved
bottom section.  Three internal stiffening ribs welded to the underside of the top plate provide stiffness and thermal
contact to the LN$_2$.

The LN$_2$ reservoir has a total capacity of 80 liters, limited
to 60 liters by the fill and vent locations and yielding
a 56 hr hold time when MMIRS is zenith pointing.  The hold time is comfortably
greater than 24 hrs in normal operation.

\subsection{Flexure performance}

The flexure between the slit and the detector is $\leq$ 1 pixel, measured between zenith angle 0 and 45$^{\circ}$, at all rotation angles. During the course of a typical 2 hour observation block we see less than 0.2 pixels flexure.  Our design goal was less stringent: $<$1 pixel in any 2 hour observation period.   The differential flexure between the guider and the slit was measured in the lab to be approximately 1 pixel between 0 and 60$^{\circ}$ zenith angle.  We are able to observe up to 2 hours following mask alignment with a misalignment of less than 25\% of the slit width. Under these conditions slit losses due to misalignment are typically less than 10\%.

\subsection{Vacuum system}

MMIRS has a built-in turbo pump system used to evacuate the slit mask
chamber and the main chamber. The entire instrument can be pumped to a
pressure of 10$^{-4}$ Torr in less than 4 hrs. Each chamber has a cold
charcoal adsorption cartridge conductively coupled to the LN$_2$ dewar
for pumping nitrogen and other noble gases when the cryostat is cold
and the turbo pump system disconnected.  The vacuum pump system
consists of a Varian Triscroll 600 pump and V-301 turbo pump mounted
to the slit mask chamber.  The main chamber is evacuated through the
gate valve.  Vacuum gauges are mounted on right angle elbows to
minimize illumination of the optics.  Redundant gauges and solenoid
valves are provided to control the backfill process for slit mask
exchange and other operations.  This backfill system is controlled by
a safety interlock system to avoid instrument damage (see section \ref{sect:interlock}).


\section{Optics Mounts}


\subsection{Design goals}

The design goal was to limit internal stress under
any operating condition to less than 400 PSI for the crystalline
materials (CaF2, BaF2, and ZnSe) elements, and less than 500 PSI for the
S-FTM16 and Infrasil elements.  The maximum anticipated stresses are during
shipping and cool down or warmup.  We adopted a 10g maximum shipping load
and a maximum 0.15 K min$^{-1}$ temperature change during cool down and
warmup.

\subsection{Spring loaded mount design}

\begin{figure}
\begin{center}
\begin{tabular}{c}
\includegraphics[width=6in,angle=0]{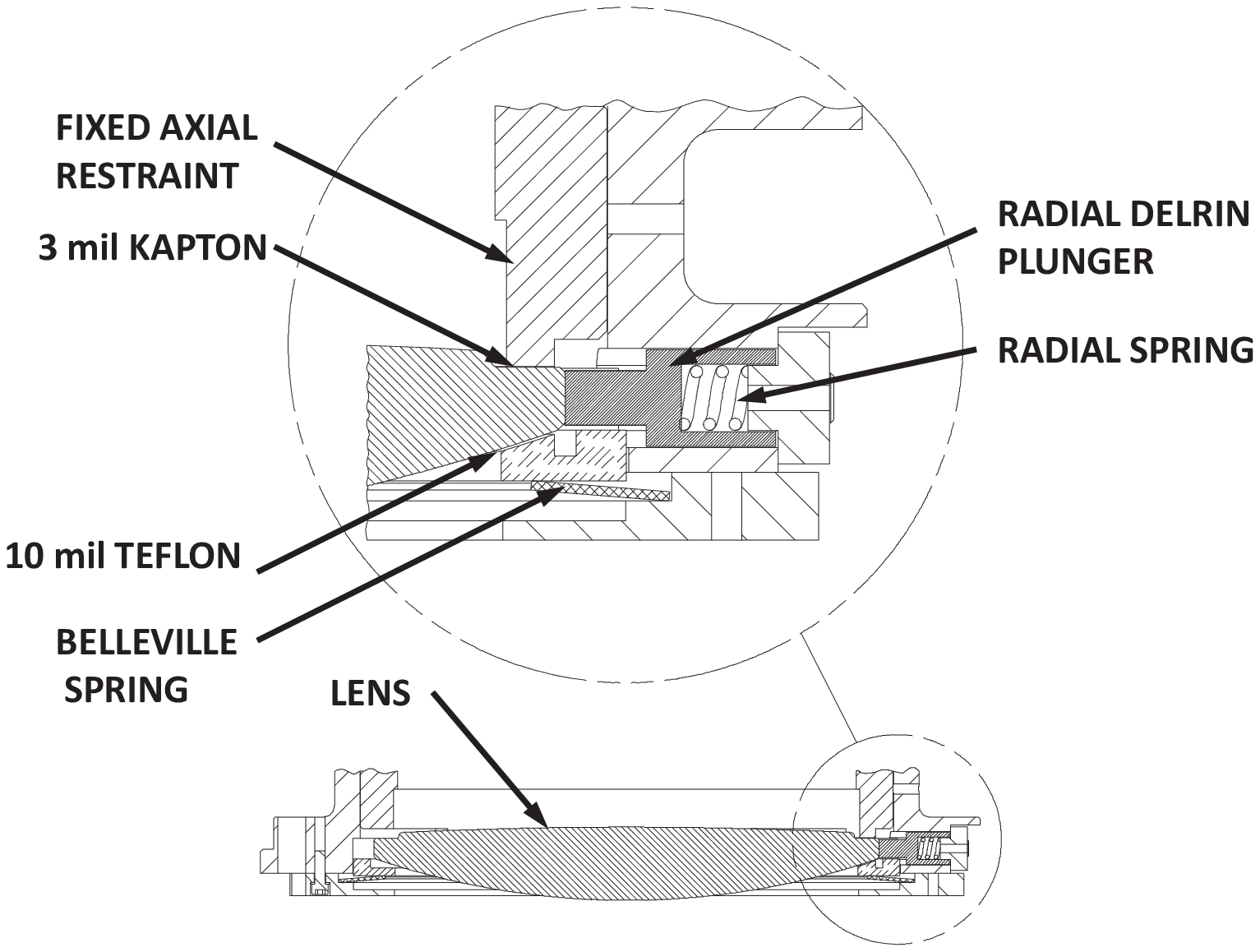}\\
\includegraphics[width=6in,angle=0]{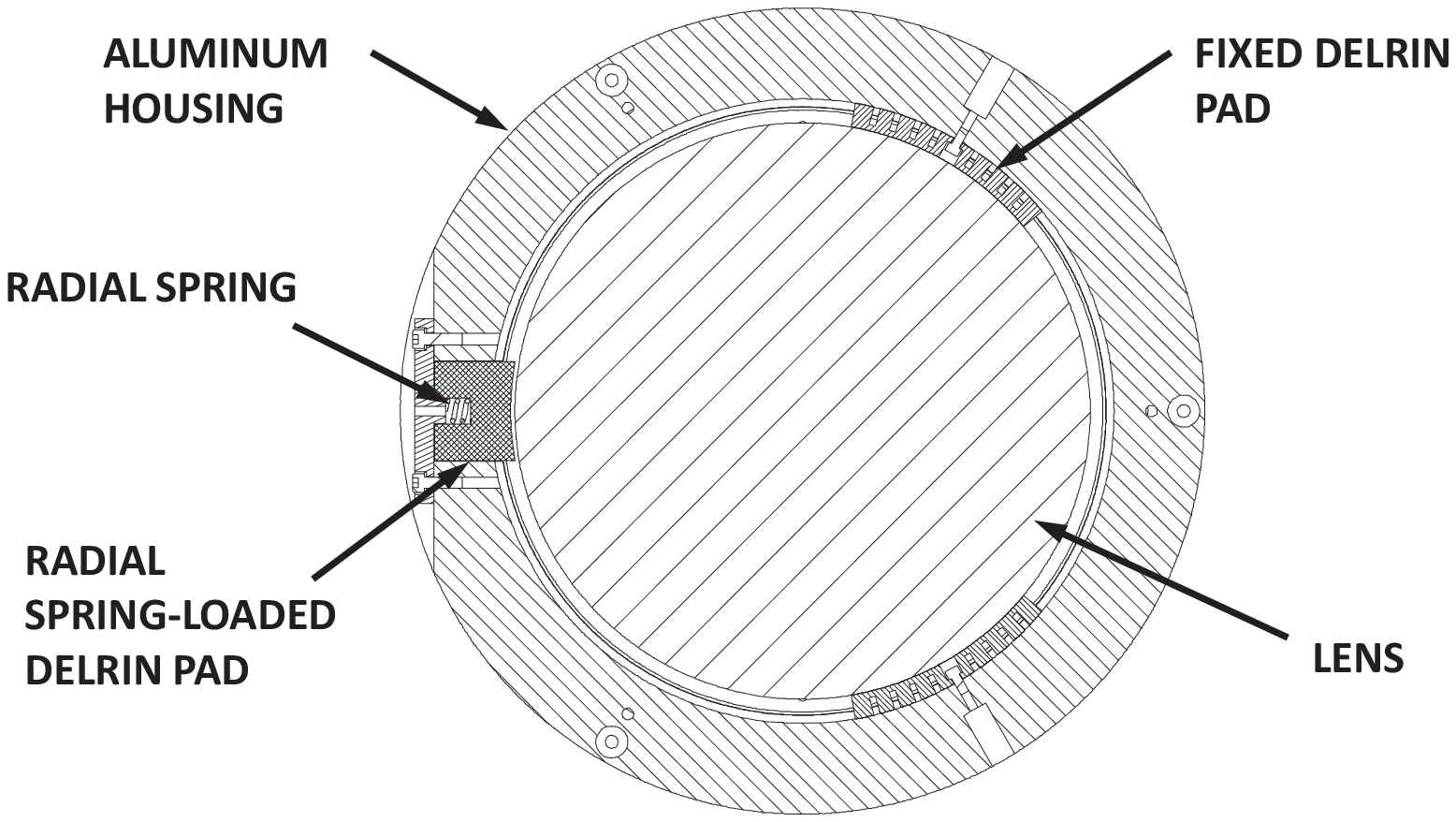}\\
\end{tabular}
\end{center}
\caption{\label{fig:lensmount} Layout of a typical lens mount showing the radial and axial supports. }
\end{figure}

The MMIRS cryogenic optics mount design (Figure \ref{fig:lensmount})
 uses aspects of the optics
mounts used in the Gemini Near Infrared Spectrograph
(GNIRS) \citep{eli06} and the Gran Telescopio Canarias' EMIR \citep{bar04}.  The lenses are mounted in
aluminum housings.  Radially they are constrained by two Delrin pads
40$^{\circ}$ in angular extent to maintain acceptable lens stresses during shipping and cool down/warmup.  The
thickness of the pads is chosen so that the lenses are centered at
room temperature and also at 77K.  All the Delrin parts were made from
the same batch of material, of which we had the CTE measured.  The radial spring
forces are provided by compression springs selected from the Lee
Spring catalog.

Axially each lens is constrained by three raised aluminum pads each 40$^{\circ}$
long and 5mm wide.  A 75$\mu$m
thick Kapton spacer is placed between the pad and lens.  The opposite side of the lens provides the axial
spring force via a Teflon spacer, an aluminum spacer with three more
matching raised pads, a custom made beryllium copper Belleville spring
and an aluminum spring retainer.  The springs were designed to provide
a 10g preload.

\begin{figure}
\begin{center}
\begin{tabular}{c}
\includegraphics[width=6in,angle=0]{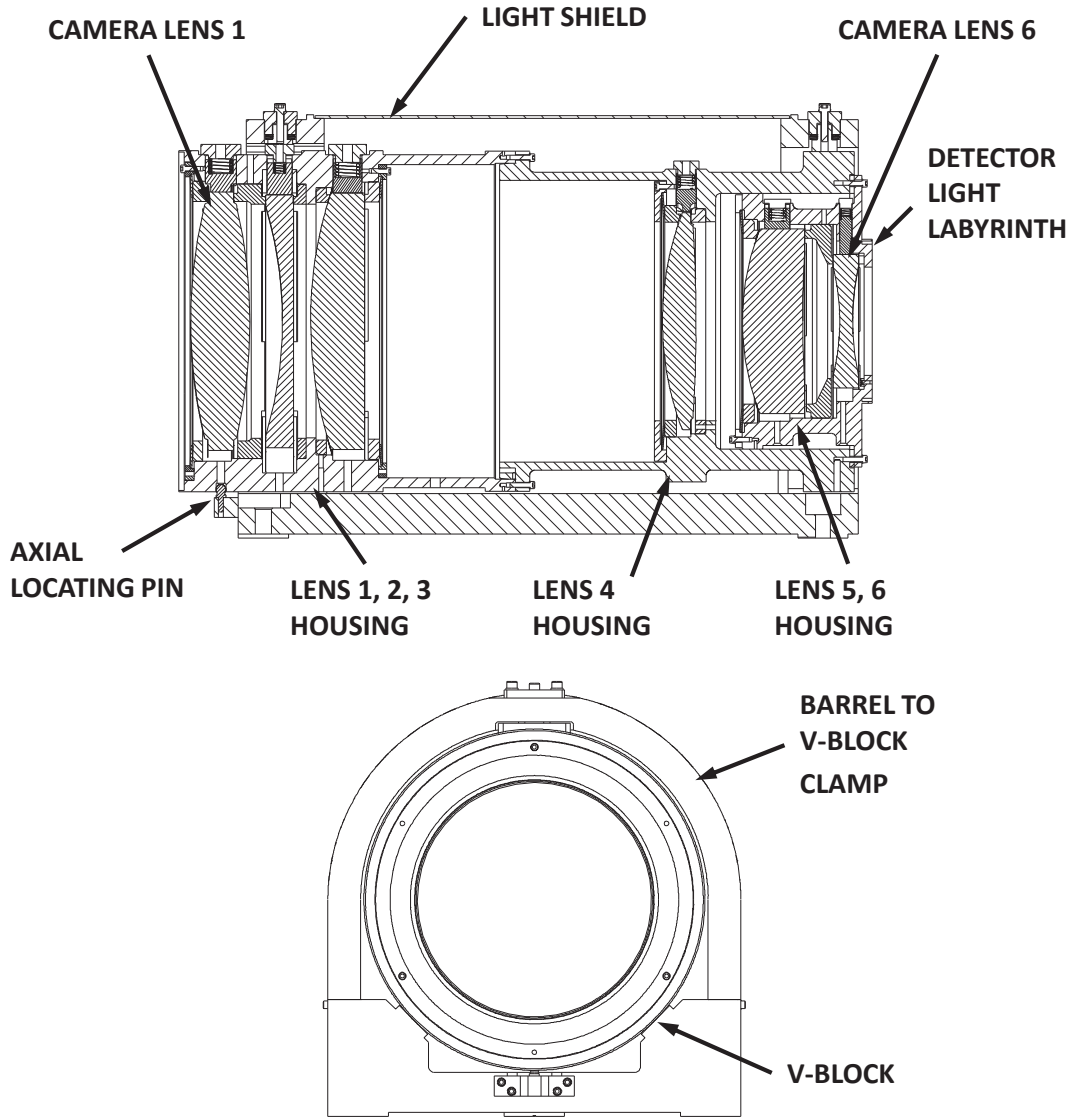}\\
\end{tabular}
\end{center}
\caption{\label{fig:camera} The camera assembly showing the overall mounting scheme for the six camera lenses. }
\end{figure}

Figure \ref{fig:camera} shows the complete camera assembly with light shield
and mounting arrangement.  The camera barrel assembled from the individual
lens mounts is clamped to a V-block base that is in turn bolted to the optical bench.
The collimator optics are assembled in a similar fashion.

\subsection{Alignment of lens assemblies}

Our tolerance analysis indicated that we can tolerate radial
positioning errors of each lens distributed uniformly within a 75
$\mu$m radius circle.  The tilt error tolerance was $<$0.1 mrad for the alignment
of each lens' optical axis relative to its optical flats, and the same
tolerance for the bezel axial flats relative to the assembly's optical axis.
No axial, radial, or tilt adjustments were provided so our approach was
to measure the lens positions as described below and to correct manufacturing
errors where possible.  Although we did not meet the tolerances above, ray tracing indicated that
our basic goal of a $<$12 $\mu$m RMS image diameter contribution from assembly errors
was met.

The process of verifying that the parts were
made correctly or correcting machining errors, developing the procedure to safely insert each lens
into its housing, and verifying that each lens ended up in the correct
place required approximately one week per lens.  Critical dimensions
of each part were measured (with digital
calipers or with a coordinate measuring machine) and then
the parts were cleaned and inspected for burrs under a microscope.

After each lens was installed we measured its decenter, tilt, and axial
placement relative to its housing using a TriOptics Opticentric
machine.  The Opticentric consists of a rotary table on an air
bearing.  The table has tip-tilt and X-Y translation manual control.
Located on either side of the table is an autocollimator equipped with
a video camera.  Using an appropriate head lens on each autocollimator
focused at the center of curvature of the lens surface we see the
autocollimator cross hairs in focus on the camera.  We then adjust the
tilt and translation of the table to bring the lens optical axis onto
the rotary axis of the air bearing.  The final step is to measure the
mechanical runout of the lens housing using a digital indicator.  We
can also measure the axial position of the lens by focusing the
autocollimator head lens onto the surface of the lens under alignment.  Our
measurement repeatability with the Opticentric is less than 10
$\mu$m of decenter, 0.1\,mrad of tilt, and 25 $\mu$m axially.
Our measurements show that we achieved a mean radial centration error
of 45 $\mu$m and a maximum error of 90 $\mu$m.  We achieved a
mean tilt error of 0.22 mrad and a maximum tilt error of 0.75 mrad.


\section{Electrical Design}


\subsection{Electrical overview}

The MMIRS electrical system is housed in two instrument mounted racks containing all
control electronics for the system.
Rack 1 contains all motion control and calibration system control
components and provides the interface to facility power and network connectivity.
Rack 2 houses temperature and vacuum control, and the science detector interface.
Temperature control of the detector is provided by a Lakeshore 321 controller, and
cryostat temperatures are monitored using two Lakeshore model 218 monitors. Chamber
warm-up and cool down control, as well as gate valve temperature control
is provided by four Omega Ethernet temperature controllers. Three Pfeiffer vacuum gauges
are controlled by a Pfeiffer controller. Vacuum pump control is provided by means of switched
power to an externally mounted scroll pump and a rack mounted Varian V301-AG turbo pump
controller.  Each rack contains a guider power supply and electronics
box and an Acromag six channel temperature monitor used to track rack internal temperatures.
Both racks are insulated and include a heat exchanger and fan connected to the facility liquid cooling loop.
A thermostat in each rack cuts off AC power should the cooling system fail.  The total power dissipation
in the two racks is 700 W.

\subsection{Motion control}

The motion control system is a compact PCI-bus version of the Delta-Tau UMAC
system.  All of the motion axes are driven by Phytron
stepper motors and drivers. The Phytron PAB93-70 drivers are mounted in three 4U 19 inch rack
mount boxes. The Phytron driver provides for intelligent setup and control of
motor run parameters and readout of motor temperature, bus voltage and current. The
setup of the drives is provided by a serial interface on each Phytron driver, which in turn
connects to a serial to Ethernet converter. The Delta-TAU system provides real-time
motor direction and control.

\subsection{Vacuum and interlock system}\label{sect:interlock}
A vacuum interlock system implemented in hardware
monitors the pressure and temperature on both sides of the gate valve.
The interlock system prevents the gate valve from being opened
when the two chambers are at different temperatures or pressures.  It also locks out
the back fill valves so the slit mask chamber cannot be raised to ambient
pressure if 1) the gate valve is open and either optics section is cold or if
2) the gate valve is closed and the slit chamber is cold.
Finally, the interlock prevents the slit chamber heaters from being turned on when the
gate valve is open.

\subsection{Detector electronics}

The camera controller and data acquisition interface are the same as those developed for
the earlier Smithsonian Widefield Infrared Camera on the MMT \citep{bro08}.
This controller is in turn derived from the CCD
controller used for all of our instruments for the past few years. The system
was developed to be simply extensible to as many as 72 readout channels running simultaneously at a
180 kHz pixel rate. Data acquisition and camera control is handled by a fiber-optically
coupled commercial interface module from EDT, which mounts on the I/O board of the
camera control unit. The critical task of taking the 32-channel video outputs from the preamplifier modules
and converting to digital information is accomplished on custom A/D boards.

The HAWAII-2 imager plugs into a Yamaichi zero-insertion force (ZIF) socket, soldered to
a 7-inch diameter cold header printed circuit board (PCB). This assembly serves the
functions of routing signals to and from the imager, lateral and axial support for the
spring-loading of the imager array against its precision locating structure, and as the main
cooling mechanism for the imager and buffer amplifiers.

The signals to and from the cold Hawaii-2 array
are carried on custom flexible printed circuits, terminating on four hermetic
connectors soldered to four external preamplifier modules spaced around an aluminum
mounting ring. This ring serves as system ground for the camera system, and is
electrically isolated from the rest of the structure and associated electrical grounds. The
rest of the camera controller and associated power supplies reference to this ground.

All signals and power connections coming from and going to the camera controller are
carried on RG-174 coax cabling to each of the four preamplifier modules. There is a
37-pin D-connector on the bottom of the backplane that supplies all required DC voltages
and also the digital drive signals for the imager multiplexer. Each 4-channel video
section also has a dedicated 25-pin D-connector on the backplane.

In the laboratory we explored a wide range of detector
settings and their impact on detector performance.  VRESET, for example,
affects both pixel well depth and dark current.  Our final choice of
operating voltages are BIASGATE = 3.3 V and VRESET = 1 V.  The resulting
detector performance (read noise, dark current, etc.) is discussed in
Section 10.


\section{Design issues and innovations}


\subsection{Enhanced slit mask wheel cooling}

The original design of MMIRS had the slit mask, Dekker, filter and
grism wheels, and focus stage each cooled via sapphire ball bearings.
Although formal calculations indicated that this would provide ample
cooling, in practice this was not borne out.  This shortcoming is
believed to be because of the point contacts between the balls and the
bearing grooves.  Because the focus stage has a linear motion of only
a few mm, adding copper straps was sufficient to solve the cooling
problem there.  The grism and filter wheels are enclosed in a box and
do not require particularly fast cooldown and warmup, so radiative
cooling is sufficient to cool them.  The slit and Dekker wheels pose
more of a challenge.  They are required to cool to within 10K of their
final temperature within a $\sim$5 hr period, and warm up similarly
fast.

Our current design uses Energy Science Laboratory's Vel-Therm wrapped
around the rotary shaft of the slit and Dekker wheels.
Vel-Therm is a velvet material made of highly conductive carbon fibers that are
epoxied into a backing material.   We performed laboratory tests of Vel-Therm
K40G-30B-M5 compressed to a thickness of 0.5\,mm and found that its thermal
conductivity was 630 W mK$^{-1}$ at 300K, 270 W mK$^{-1}$ at 135K, and
170 W mK$^{-1}$ at 90K.

The fibers of the Vel-Therm are canted downwards along the axis of the
shaft, allowing them to bend but not buckle as the wheel changes direction.
Teflon seals above and below the Vel-Therm trap any Vel-Therm fibers that may
come loose from the backing material.  With a Vel-Therm area of 18 cm$^2$, and
a slit wheel mass of 6 kg, we measured in the lab a wheel cool down time of 4 hrs
to cool within 20 K of the equilibrium temperature (85K) and 6 hours to cool
within 10K.  No temperature sensor is mounted on the wheel during regular
operation.  In normal operations we allot 5 hours for the wheel cool down.

\subsection{Gate valve seal heaters}

During preship testing we found that the gate valve once failed to seal properly
when it was closed.  The leak stopped during the instrument warmup, leading us
to conclude that the gate valve blade had cooled by radiation, causing the Viton
seal on the perimeter of the blade to harden.  To solve this problem we epoxied a
6 W Kapton heater pad to the valve blade, along with a temperature sensor.
When the instrument is cold we maintain the blade at -5 C and warm it up to 15C
when we exchange masks.  Since the addition of the heater we have had no problems
with gate valve leaks.

\subsection{Gate valve thermal glow and enhanced shielding}

The gate valve separating the slit mask chamber from the main chamber makes it practical
to change slit masks on a daily cycle but its seal must be near room temperature to seal properly.
Because the warm gate valve is located between the cold slit mask and the cold collimator optics
it must be carefully shielded to avoid
intense thermal background through any filter that passes K band light.  We designed a cold
spring-loaded baffle traveling on THK rails that passes through the open gate valve
and nests into a fixed cold baffle above the first collimator lens.
A cam on the gate valve blade raises the baffle above the gate valve
as the valve is closed.  Designing a smoothly operating baffle mechanism with sufficient
travel to seal tightly took several iterations.  For future infrared spectrograph designs
we recommend using a cassette of cold slit masks that are admitted through a gate valve
above the telescope focus.  This arrangement eliminates all warm surfaces below the slit \citep[e.g.][]{fab12}.

\subsection{Warm up heater adhesion failure}

The original warmup heaters for MMIRS were Minco foil heaters epoxied to the curved surfaces of the two LN$_2$ vessels with Lord 3170 epoxy. After the first warmup of the slit mask dewar from 77K to room temperature, we found that 8 of the 12 heaters were debonded from the vessel and melted.  Rather than conduct an extensive investigation and requalification program for this type of heater, we switched to cartridge heaters soldered into copper blocks and bolted to the top surfaces of the two dewars.  These heaters have performed without any problem.

\subsection{Coating failure}

After $\sim$ 2 yrs of operation we noticed that the antireflection coating on the inner surface of the
second corrector lens had failed, becoming soft and iridescent with reduced throughput.
This lens is cooled and warmed up whenever the slit masks are replaced, so we
suspect a manufacturing issue exacerbated by thermal cycling. We manufactured and installed a replacement lens that
is showing no degradation after 1 yr of operation.

\subsection{Intermittent Hawaii-II electronics}

The MMIRS detector is a first-generation 2048$\times$2048 pixel HAWAII-2 detector and unfortunately one of its 32 output channels is now failing.   When the output fails, 6\% of the spectra are lost.   The failure is intermittent, with an average failure rate of 10\%, ranging on different nights between essentially zero frames up to 25-35\%.  This intermittent failure could be intrinsic to the array or could be associated with the readout electronics and cabling.  The intermittent nature of the failure and the cryogenic environment make the problem difficult to diagnose.

In the years since we purchased the HAWAII-2, IR array technology and the associated readout electronics have progressed significantly.  Current HAWAII-2RG arrays have greater reliability, higher and more uniform quantum efficiency, greater bias stability, better linearity, reduced image persistence, and an order of magnitude lower dark current than our installed array.    Further, application-specific integrated circuits (ASICs) are available for the new arrays to provide simpler, high-reliability readout electronics with much simpler cabling.  We have begun an $\sim$1 yr long effort to purchase and install a new HAWAII2-RG array in MMIRS, and hope to complete this process in the fall of 2013.


\section{Software}


\subsection{Overview}

The MMIRS software supports slit mask design, instrument preparation procedures, night-time observing, and pipeline data reduction.  The majority of MMIRS software is written in Tcl with some low-level code written in C.  The system's client/server architecture is implemented using a simple ASCII over TCP sockets protocol.  Typically, a server controls each hardware subsystem.  Server commands expose the hardware functions to the user interface and to script clients that are used to operate the instrument.  To minimize external dependencies, all interfaces with the Magellan or MMT facilities are localized in a single site specific telescope server component.  The MMIRS observing manual contains detailed instructions on running the slit mask design and observing software.

\subsection{Slit mask design software}

MMIRS slit masks are laser cut into black-anodized aluminum blanks 75-200 $\mu$m thick. The observer downloads and runs the slit mask design program well in advance of a run to generate the mask files that are later converted into laser cutter commands. The observer uses a graphical user interface (GUI) to enter input parameters and the name of a catalog of observing targets.  The catalog identifies targets with their priorities, mask alignment stars, and the minimum and maximum number of slits allocated for each target priority.  A set of disperser/filter combinations is chosen and the mask design is valid for observation with the specified combinations.  A field center or list of field centers for each proposed mask is provided and the software will attempt to optimize the number of slits fit by varying the mask position angle.  If multiple masks are fit at the same time the distribution of targets on each mask will be optimized between masks.  The user interface retrieves candidate wavefront sensor and guide stars from the GSC2 catalog and the availability of appropriate stars is verified for each mask position.

Following mask optimization the user is presented with a list of mask designs ranked by the number of targets fit weighted by user priority.   Selecting a mask design from this list displays the mask overlaid on a FITS image of the field.  The overlay marks targets, slits, guider and wavefront sensor regions, and candidate guide stars.  A second display shows dispersed spectral regions superposed on a graphic of the detector array.
Two text windows display the target catalog and the mask design with slits and targets.  Selecting a line in either window highlights the corresponding target or slit in the graphical displays.  After reviewing the possible mask designs the user selects the final mask files to be forwarded for fabrication.

\subsection{Instrument preparation software}

The MMIRS instrument specialists on the Magellan staff oversee thermal cycling of MMIRS through a dedicated user interface.  The instrument specialist selects the desired operation: 1) initial cool down of the slit mask and camera sections at the start of a run, 2) slit mask exchange or 3) warmup of both instrument sections at the end of a run.  The instrument specialist executes each step of the procedure as directed by the program.  The software checks that temperature, pressure, and motion stage positions are correct before proceeding. If a procedure is blocked, a text message directs the instrument specialist to address the listed issues.

\subsection{Wavefront sensor and guide star acquisition software}

The telescope operator interacts with the wavefront sensor/guide star acquisition interface to choose appropriate stars for use in both guide cameras.  The operator is presented with a star chart showing the stars of appropriate magnitude available in each of the guide camera patrol regions.  Selecting a star moves the camera to that star and allows it to be centered in the wavefront sensor aperture or guide camera.

\subsection{Slit mask and long slit alignment software}

After the telescope operator has acquired the guide stars, the observer must align the mask or long slit prior to observing.  This procedure is carried out by following a sequence of steps presented to the observer in the mask alignment interface.  The mask alignment software measures the relative offset between the mask alignment stars and the centers of the alignment boxes cut in the mask using a pair of science detector images.  If the field is not well enough aligned to allow the alignment stars to be viewed through their boxes, the images are taken with the slit mask removed and the box positions are estimated from the nominal positions in the mask design file.  A least squares fit of the position errors of the alignment stars is presented to the user and stars with large errors can be discarded.  When a good fit is achieved, the computed position error is sent to the telescope as a guide correction.  This process is iterated until the mask is properly aligned on the sky.  The alignment software also creates a telescope offset catalog that is used to successively place a telluric calibration star in each of several slits.  A similar procedure is followed to align long slits.  Alignment of a slit mask typically takes 5-10 minutes.

\subsection{Observing interface}

The observing interface allows the user to start MMIRS from a complete shutdown and to acquire science images in either manual mode or observing catalog mode.  In manual mode, the observation is fully specified from the interface and the user presses ``Go" to take a single exposure.  In catalog mode, an input catalog specifies a subset of the observing parameters, e.g. filter, mask, exposure time, etc.  Catalog-specified parameters are disabled on the interface.  When ``Go" is pressed an exposure is taken for each row in the observing catalog.  Prepared catalogs are available to automate calibration sequences and dark exposures.
An instrument status window displays a schematic drawing of the MMIRS' current mechanical state, showing the positions of the motion stages and the current exposure status.

\subsection{Data archive}

Three copies of the original data are made for archival purposes.  After each exposure the data are copied to an observer directory for analysis and observer transport to their home institutions.  Each day the LCO staff makes a second copy for the archive in La Serena.  Finally, a third copy to an external hard drive is prepared for shipment to the CfA where the data are imported into a permanent online archive.

\subsection{Engineering software}

Additional engineering software interfaces allow troubleshooting of the motion control and cryogenic/vacuum systems.  Access to low level motor control and status as well as individual temperature and pressure sensors are available from these interfaces.  These interfaces are not intended for use during normal observing procedures.

\subsection{Data reduction pipeline}

\subsubsection{Pipeline overview}

The MMIRS long slit and multi-slit data reduction pipeline is a stand-alone package implemented in
{\sc idl} with the first step implemented in {\sc C++}. Complete details
will be provided in a future paper (Chilingarian et al. in prep); here we give an
overview of the pipeline.

The pipeline is controlled by a task control file having a format similar to
FITS headers, i.e. a set of keyword -- value pairs where the keyword length
is limited to 8 characters and the total line length is limited to 80
characters. The control file can be created either in a text editor or by an
automated system that analyzes observing logs.
The control file can be edited in order to perform specific data reduction steps
but by default the data are reduced completely until the sky subtraction
step.

\subsubsection{Primary data reduction and wavelength calibration}

The primary data reduction includes the following steps:
\begin{itemize}

\item \emph{Fitting the up-the-ramp slope} in every pixel to produce a
2-dimensional image for each exposure. This reduction is performed by the {\sc
mmfixen} package implemented in {\sc C++}. {\sc
mmfixen} also corrects for detector nonlinearity and saturation.
Hot pixels and pixels saturated in a single read are masked at this
step.

\item \emph{Dark subtraction} uses an average of up-the-ramp processed 2D frames.
We typically average 5 dark frames obtained for every readout interval and exposure time.
We use different average dark frames for science exposures, flat fields, comparison spectra
and telluric standards because different exposure times and readout intervals are used.

\end{itemize}

After the primary data reduction is complete, we subtract pairs of dithered
spectral exposures (difference images hereafter) and bring them through the
rest of the pipeline steps along with the individual exposures. We retain
individual exposures to correctly propagate photon statistics to the final data products.

\begin{itemize}

\item \emph{Scattered light subtraction} is performed only for slit mask
observations and is unnecessary for subtracted pairs of
dithered spectra. The pipeline analyzes the mask definition files, then
uses them to trace the spectra and gaps between spectra on a dark-subtracted
flat field exposure. A model for scattered light is constructed
from gap counts in all types of frames (science, flat, comparison spectra,
telluric standards) using low-order polynomials along the dispersion direction
and basic splines ($b$-splines hereafter) in the perpendicular directon.
We cannot correct
the ghost spectra arising from reflections from the mask back surface
in this fashion.
However, the amplitude of ghost images is about 1\% of the flux
in the corresponding slits and can be ignored unless very bright sources were observed or used as
alignment stars. With long slits there are no gaps to allow
modeling scattered light.

\item \emph{Mapping of optical distortions} uses the spectral
traces from the previous step. We use a low order 2D polynomial
approximation to describe the distortion introduced by the optics and
misalignment of the disperser relative to the detector array.

\item \emph{Extraction of two-dimensional spectra} is performed only for
slit mask data. The pipeline uses the mask definition files to extract 2D
spectra for each slit from the original frames without resampling. This
operation is performed on both individual
exposures and subtracted pairs.

\item \emph{Flat fielding} is an important step in the data reduction
as pixel-to-pixel variations in the detector reach
35\%. Flat fielding also compensates for slit imperfections.  We normalize the flux to
unity using the 5$\times$5 median-averaged frame as a reference to
handle cosmic ray hits or hot pixel residuals. For slit mask data we
normalize to the maximal flux within each slit to accommodate the intensity
variations in the internal flat field calibrations.

\item \emph{Wavelength calibration} differs for long slit and
slit mask data: the fit uses the entire frame for long slit data and
extracted 2D frames for slit mask data.
First, the Ar comparison lines are identified and an
initial wavelength solution is fit with a 2D third order polynomial.
Initial wavelength calibration of alignment stars uses an approximate 1D dispersion
relation because the calibration lines are blended due to the large alignment star
apertures. The final wavelength solution uses OH airglow lines to refine the
initial wavelength solution. A template spectrum for each aperture is extracted for
a single pixel close to the frame center in the cross-dispersion direction.
Spectra from adjacent pixels are cross-correlated with the template spectrum
to measure and apply small residual wavelength shifts.
This approach may fail when bright targets reduce the correlation coefficient
between the template and adjacent pixels. In this case the same
procedure can be performed on the comparison spectrum with a minor loss in accuracy.

\end{itemize}

\subsubsection{Sky subtraction}

The subtraction of night sky emission is one of the most critical steps in
the near infrared data reduction. The night sky emission includes bright OH emission
lines and a continuum background that vary with time and pointing position.
Faint targets observed with MMIRS can be hundreds of times fainter than the
night sky level, so careful estimation of the sky background is essential.

In the MMIRS pipeline we use a hybrid approach to sky subtraction.
The difference images that we create early in the pipeline are the
first step in sky subtraction.  However, the difference images contain residual night sky
emission originating mostly from temporal variation of OH line flux.
For a typical 300 s exposure these residuals may reach a few percent so
we apply a slightly modified version of the \cite{kel03} technique to the
difference images. This technique uses
spectra that have not been resampled and precise pixel-to-wavelength mapping to
create an oversampled model of the night sky spectrum that is
parameterized with $b$-splines and evaluated at every position at every
slit. This techniques avoids artifacts from interpolation of
undersampled night sky lines.

This Kelson approach works well on long slit data where
the slight shift in dispersion for off-axis rays allow good sampling of the emission line
shape.  For multiple slit data the short slit length doesn't allow full sampling.
We used a modified approach creating the sky model from all of the slitlets
assuming that their widths are equal.  We must accommodate
residual flat fielding errors along and perpendicular to the dispersion.
We use a 3D $b$-spline/polynomial
parametrization where the $b$-splines are fit in the dispersion direction while
2D Legendre polynomials that depend on X and Y slit positions
account for the residual flat fielding errors.
Our modified approach applied to the difference images leaves OH line residuals
consistent with the Poisson photon statistics.

\subsubsection{Final cosmic ray rejection, wavelength calibration, and rectification}

After sky subtraction we perform the
final cosmic ray and hot pixel cleaning using the Laplacian
filtering technique \citep{van01} on difference images. We modified the
original code to handle negative and positive features
because we are dealing with difference images.
Wavelength calibration and geometric rectification are then applied
using the distortion map and wavelength solutions derived in
previous steps.

\subsubsection{Telluric correction}

Telluric absorption correction is necessary to remove the variable OH absorption
bands that punctuate near infrared spectra.
We correct for this absorption by
observing telluric standard stars, typically A0V stars with relatively featureless
spectra observed at airmasses bracketing the science
observations \citep{vac03}. The observed spectra are compared with model predictions to derive
a transmission curve. This procedure allows relative flux calibration
of the data.

\subsubsection{Final data products}

The pipeline steps above produce calibrated 2D difference images.
A final 2D spectrum is produced by inverting and adding the negative
portion of the difference image. Finally, we extract 1D spectra using optimal extraction
with a double-Gaussian profile derived from alignment stars.
The final data products are available in several formats: 1) multi-extension
FITS files with one extension for each 2D co-added spectrum, 2)
single-extension FITS files with all extracted 1D spectra, 3) a
Euro3D-FITS file \citep{kis04} for 2D extracted spectra, and 4) a Euro3D-FITS file for 1D
extracted spectra. The first two files are accompanied by files in the
same format providing flux uncertainties, while this information is stored
internally in the Euro3D-FITS format. MMIRS slit mask spectra in the Euro3D-FITS format
contain metadata making them Virtual Observatory compliant \citep{chi06}
and easy to visualize with the tools described in \cite{chi08}.


\section{Daytime operations}


\subsection{Slit mask exchange procedure}

The MMIRS slit mask wheel holds 9 slit masks that are typically exchanged during
day time by mountain staff.  The procedure takes 10-11 hrs: 5 hrs for warm up,
1 hr for mask exchange, and 4-5 hrs for cool down.  A step by step software
procedure guides the mask exchange operator through the process.  The interlock
system is an additional safety mechanism to prevent damage to the optics and
detector.

While the slit mask chamber is warming up, the slit mask operator uses a pull down
menu to select the masks to be installed in each of the 9 slots.  When the process is
complete, the mask configuration is saved and an observing catalog and mask exchange
plan printout is created.  After the slit masks are installed and cooled down to operating
temperature, the masks are illuminated and a FITS reference file for each mask is
recorded.

\subsection{Controlled cool down}

The cool down and warmup of the optics section of MMIRS is controlled
to minimize stress on the optics and detector.  The cool-down control is
accomplished with a Valcor cryogenic valve, an Omega
temperature controller, and an RTD mounted on the curved back surface
of the dewar at the end under the detector.  The set point is ramped
downwards at 0.15K min$^{-1}$ through software.  A threshold prevents the
set point from dropping more than 5 K below the actual temperature.
This is a safety feature so that if the flow of cryogen is temporarily
interrupted and then restored (e.g. when swapping from an empty LN$_2$
tank to a full one) there will not be a sudden drop in temperature.
A 1.6\,mm orifice at the end of the LN$_2$ fill line prevents the
maximum flow rate of LN$_2$ from exceeding the nominal flow rate
by more than a factor of 2, guarding against rapid cooling in the
unlikely event that the valve sticks open.

The warmup is controlled in a similar manner.  Cartridge
heaters soldered into a copper block are bolted to the top of the
optical bench.  Additional low power heaters are mounted onto the detector
board.  The detector board heaters, driven by a Lakeshore
controller, are used to stabilize the temperature during normal
operation.  During warmup the detector board is ramped up at 0.15K/min 
and the optical bench heater set point is set a
few degrees below that of the detector board to minimize freezing
outgassed material onto the detector.  Again thresholds prevent the
set points from deviating from the actual temperatures so that loss of
power will not cause a sudden drop in temperature.

Thermostats mounted on the heater blocks prevent current from going
through the heater when the temperature rises above 40 $^{\circ}$C.
The total heater power is limited to a level twice the nominal requirement. These features guard
against catastrophic heating in the event of a control system failure.


\section{Observing Protocols}


\subsection{Introduction}

For all observing modes, dark current is estimated from dark frames taken at
the end of every night. A script scans all the observations
carried out during the night and compiles a list of dark exposures
available for data reduction.

\subsection{Imaging}

Imaging exposures normally use a
pseudo-random dithering pattern within a specified square region larger than the
target's extent, typically
30$^{\prime\prime}$$\times$30$^{\prime\prime}$ to 180$^{\prime\prime}$$\times$180$^{\prime\prime}$.
Observing dithering catalogs including 100
pseudo-random positions are calculated by the observing software once the
observer specifies the exposure time. Exposure times should not
exceed 30~sec in the $K$ band, 60~sec in the $J$ and $H$ bands, and 120~sec
in the $Y$ band in order to avoid saturation by the sky background and
bright field stars.

Twilight and internal flat field images should be acquired
during the same night or at least during the same observing run. Normally,
there is no need to take photometric standards as MMIRS' large
field of view contains numerous 2MASS sources that supply
absolute photometric calibration.

\subsection{Long-slit spectroscopy}

Following alignment we recommend dithering the target between
3 to 5 equally separated positions along the slit. The separation between dithering
 positions should normally exceed
the target spatial extent and is typically 10-120$^{\prime\prime}$.
Exposure times depend on the source and typically range
from 5-10~s with 1 second readout intervals for bright stars to 600~s with 10 s
readout intervals for faint targets. We recommend 300 s exposures with 5 s readout intervals
for faint targets.

Long slit observations require the following
calibrations: 1) Ar comparison spectra and internal flat field frames at least once
for each target and at 2 hr intervals for long integrations and 2) multiple telluric
standard stars (A0V spectral type) observed at the range of airmasses encountered.

\subsection{Multi-object spectroscopy}

Following alignment we recommend dithering the target along the slit as described above.
The optimal dithering spacings are a multiple of 0\farcs2 to shift by an integer
number of pixels. Two position dithering is
usually selected for bright targets while four position dithering can be
used to improve sky subtraction. The standard slit height is 6-7$^{\prime\prime}$
so typical choices are $\pm$2$^{\prime\prime}$ or
$\pm$1.8$^{\prime\prime}$ and $\pm$1.4$^{\prime\prime}$ where the reference is
centered along the slits.  The recommended exposure strategy is the same as that used
for long slit observations.

We recommend the same sequence of calibration exposures as for long slits
described above.  The Ar comparison and internal flats should be obtained
before moving the mask wheel because the repeatability of the wheel position
is not perfect.  A script aids in acquiring telluric standards by
placing the standard star into every slit or at least into 5 representative
slits: central, leftmost, rightmost, lowermost, and uppermost.  The procedure
covers the maximum wavelength range and assesses
transmission variations across the field of view.


\section{Scientific performance}


\subsection{HAWAII-2 array performance}

\subsubsection{Readout scheme and readout noise}

Our standard readout mode for exposures longer than 30 seconds is "up-the-ramp", where we read out the detector non-destructively every 5 s.  We perform a linear least squares fit to the series of readouts to determine the count rate for each pixel, rejecting anomalous intervals affected by cosmic ray hits.  Up-ramp-mode also allows increased dynamic range because data from pixels which saturate part-way through the exposure can still be used.  In a typical 300 second spectroscopic exposure the read noise is reduced from 17 $e^-$ for a double-correlated read to 5 $e^-$ for up-the-ramp sampling.  Better performance could be obtained with a HAWAII 2RG array and a by increasing the number of readouts.  Table \ref{table:array} summarizes the performance of our HAWAII 2 array with our current readout scheme.

\subsubsection{Dark current}

The dark current in the detector currently dominates the noise performance for J and H band spectra. The current median dark current is 0.1 to 0.3 $e^-$ pix$^{-1}$ s$^{-1}$ at 78 K depending on the recent illumination of the detector. The dark current is higher in the corners of the device. The lowest dark current measured in the laboratory was 0.08 $e^-$ pix$^{-1}$ s$^{-1}$ at 80.5K, which after compensating for the temperature difference is a factor of two lower than the current value. The lower dark current was achieved with a VRESET=0.75 V instead of the current 1.0 V. Lowering VRESET to 0.75 V reduces the full well capacity by 25\%. The dark current increases after the detector is exposed to light; this effect is known as image persistence. For example, the dark current will be higher for many minutes after taking a mask alignment image.  Replacing the HAWAII 2 detector with a HAWAII 2RG will decrease the dark current by an order of magnitude.

\subsubsection{Linearity and full well depth}

The electronics gain is set at 5 $e^-$ DN$^{-1}$.
Our choice of VRESET provides a pixel full well depth of 230,000
e$^-$ or 46,000 DN. Tests indicate that the array is
linear at the 1\% level until counts reach 90\% of full well depth.

\subsubsection{Reset anomaly}

A reset anomaly is commonly found in HAWAII-2 detectors, including ours.
After clearing the detector, the first read has
systematically higher counts than subsequent reads in an up-the-ramp dark
exposure.  The reset anomaly is repeatable, but we chose to minimize its
effect by clearing the detector on a 5 s cadence between exposures.

\subsection{Backgrounds and sensitivities}

\subsubsection{Interline background}

The highest signal to noise regions of infrared spectra are found between the numerous OH air glow lines.  After accounting for the known sources of
internal backgrounds we can measure the intrinsic interline emission level. The best available estimate of the true interline background is the measurement by \cite{mai93} of 590 photons m$^{-2}$ s$^{-1}$ $\mu$m$^{-1}$.  We measure an H band typical interline background that is 25\% higher than the \cite{mai93} value with no moon illumination.  Our measurement is consistent with recent measurements with very different spectrographs including \cite{ell12} through OH suppression
fibers and \cite{sul12} with a cross-dispersed echelle spectrograph.  We refer the reader to the interesting discussions about the possible origin of this interline background in these two recent papers.

\subsubsection{Thermal emission from gate valve}

Emission from warm surfaces in the gate valve provides an extra source
of background in HK spectra. Improving the gate valve baffle has
significantly reduced this background and has decreased the affected
fraction of the detector. Tests performed with the K$_s$ filter show
that this background will be reduced below the level of the interline
background with a new HK filter that cuts off at 2.35 $\mu$m, instead
of the present 2.49 $\mu$m.  The existing HK filter will remain
available for those projects requiring coverage out to 2.49 $\mu$m.

\subsubsection{Measured throughput and sensitivities}

Table \ref{table:imageprop} give the measured system throughputs and sky count rates for imaging.
Throughput numbers were calculated by aperture photometry of a star in a 4$^{\prime\prime}$ diameter aperture and correcting for estimated extinction at 1 airmass.  The throughput includes telescope reflectivity, transmission of the MMIRS optics and filters, as well as detector quantum efficiency.
Introducing a disperser lowers throughput by $\sim$30\%.
Measured imaging and spectroscopic sensitivities are given in Table \ref{table:imagesens} and Table \ref{table:specsens}.

\subsection{Sample spectra}

\begin{figure}
\begin{center}
\begin{tabular}{c}
\includegraphics[width=6in,angle=0]{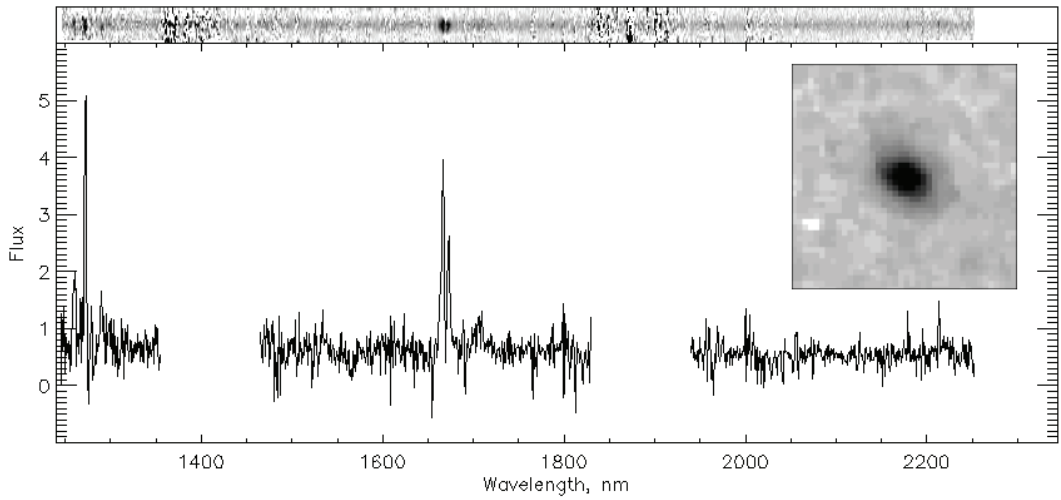}\\
\end{tabular}
\end{center}
\caption{\label{fig:spec1}  Spectrum of an H=19.9 (Vega, H$_{AB}$=21.3) Seyfert-2 galaxy at  z=1.544.  [OIII] is visible at the blue end and H$\alpha$+[NII] is visible near the center of the spectrum.  [SII] is the small bump at ~1700 nm.  The excellent OH sky line rejection of MMIRS' pipeline is apparent in this deep 7 hour exposure.  The exposure time would be considerably reduced with a narrower slit (1$^{\prime\prime}$ slit used here), the higher-dispersion H grism, and the new HAWAII-2RG array.   An HST image of the galaxy is shown in the inset - the scale is 2.5$^{\prime\prime}$ on a side.  A 2D image of the sky subtracted spectrum is visible at the top of the image. }
\end{figure}

\begin{figure}
\begin{center}
\begin{tabular}{c}
\includegraphics[width=6in,angle=0]{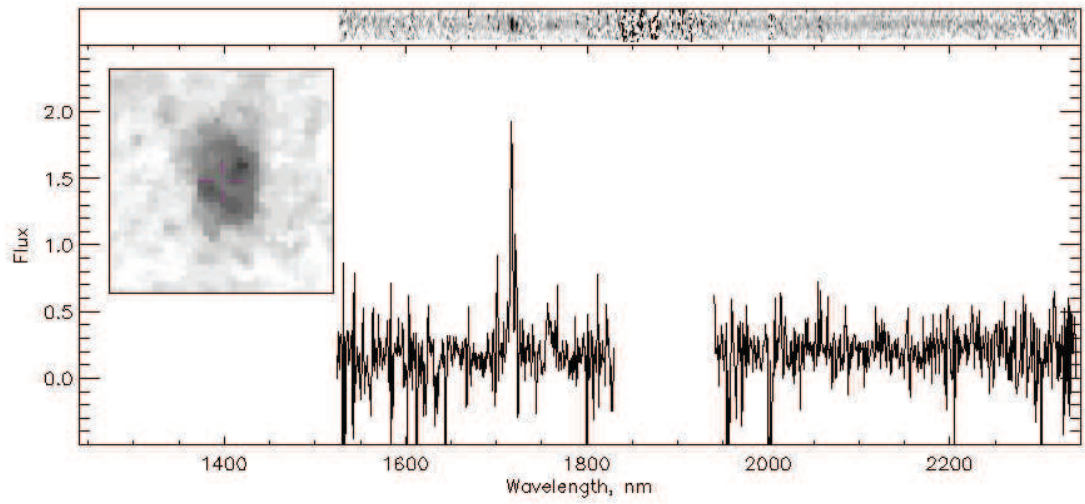}\\
\end{tabular}\end{center}
\caption{\label{fig:spec2} Spectrum of an H=20.9 (Vega, H$_{AB}$=22.3) star-forming galaxy at z=1.613.  As in the upper spectrum, H$\alpha$+[NII] is visible near the center of the spectrum, as well as a faint [SII] feature.  The galaxies in Figures \ref{fig:spec1} and \ref{fig:spec2} were observed simultaneously; the different spectral coverage is due to the placement of the slit on the mask. }
\end{figure}

We conclude by displaying spectra of two faint galaxies that were acquired on the same aperture plate \citep{lei12}.  These deep spectra of H=19.9 and H=20.9 galaxies displayed in Figures \ref{fig:spec1} and \ref{fig:spec2} demonstrate the performance of the instrument and pipeline reduction.

\section{ACKNOWLEDGMENTS}

This material is based upon work supported by AURA through the
National Science Foundation under Scientific Program Order No. 5 as
issued for support of the Telescope Systems Instrumentation Program
(TSIP), in accordance with Proposal No. AST-0335461 submitted by AURA.

We thank the staffs of the MMT and Las Campanas Observatories for their
tireless help during MMIRS commissioning.  We thank Andrew Szentgyorgyi
for his dedication and leadership commissioning the f/5 optics at Magellan.
We thank S. Eikenberry, N. Raines, and J. Julian at the University of
Florida for their support in making the FLAMINGOS 2 design available
to the MMIRS design team.


\clearpage

\clearpage

\clearpage

\clearpage

\clearpage

\clearpage

\clearpage

\clearpage

\clearpage

\clearpage

\clearpage

\clearpage

\begin{deluxetable}{rlrrrrrl}
\tabletypesize{\scriptsize}
\tablecaption{As-built optical prescription\label{table:prescrip}}
\tablewidth{0pt}
\tablehead{
\colhead{Surface} & \colhead{Surface Type} & \colhead{Radius} & \colhead{Thickness} & \colhead{Glass} &
\colhead{Diameter} & \colhead{Conic} & \colhead{Comment}
}
\startdata
 OBJ & STANDARD  &    Infinity  &    Infinity  &    AIR\_10C\_554T  &          0  &           0 & \\
 STO & STANDARD  &    -16255.3  &           0  &            MIRROR  &     6502.4  &          -1 & \\
   2 & STANDARD  &    Infinity  &   -6179.235  &    AIR\_10C\_554T  &          -  &           0 & \\
   3 & STANDARD  &   -5150.890  &           0  &            MIRROR  &   1592.584  &     -2.6947 & \\
   4 & STANDARD  &    Infinity  &    6179.235  &    AIR\_10C\_554T  &          -  &           0 & \\
   5 & STANDARD  &    Infinity  &    1430.149  &    AIR\_10C\_554T  &          -  &           0 & \\
   6 & STANDARD  &    2209.039  &      50.040  &    CAF2\_10C\_ABS  &     234.15  &           0 & Lens 1 Corrector\\
   7 & STANDARD  &   -1492.450  &      20.000  &                    &     234.15  &           0 & \\
   8 & STANDARD  &     229.065  &      20.090  &     CAF2\_4C\_ABS  &     207.35  &           0 & Lens 2 Corrector\\
   9 & STANDARD  &     162.343  &     310.000  &                    &     207.35  &           0 & \\
  10 & STANDARD  &    Infinity  &      99.398  &                    &         -   &           0 & Focus\\
  11 & EVENASPH$^a$ &  235.280  &      50.659  & CAF2\_76K\_CHARMS  &     149.31  &           0 & Lens 1 Collimator\\
  12 & STANDARD  &    -165.238  &     146.536  &                    &     149.31  &           0 & \\
  13 & STANDARD  &     -86.493  &       6.092  & CAF2\_76K\_CHARMS  &     100.33  &           0 & Lens 2 Collimator\\
  14 & STANDARD  &     104.594  &      89.332  &                    &     100.33  &           0 & \\
  15 & STANDARD  &    Infinity  &      22.872  &  BAF\_76K\_CHARMS  &     123.20  &           0 & Lens 3 Collimator\\
  16 & STANDARD  &    -104.278  &       7.234  &                    &     123.20  &           0 & \\
  17 & STANDARD  &    -121.250  &       7.816  & ZNSE\_76K\_CHARMS  &     134.12  &           0 & Lens 4 Collimator\\
  18 & STANDARD  &    -135.736  &      11.041  &                    &     134.12  &           0 & \\
  19 & STANDARD  &    Infinity  &      11.069  & I301\_76K\_CHARMS  &     141.54  &           0 & Lens 5 Collimator\\
  20 & STANDARD  &     406.740  &      29.090  &                    &     141.54  &           0 & \\
  21 & STANDARD  &    1291.376  &      20.586  & CAF2\_76K\_CHARMS  &     142.72  &           0 & Lens 6 Collimator\\
  22 & STANDARD  &    -220.874  &      50.000  &                    &     142.72  &           0 & \\
  23 & STANDARD  &    Infinity  &      10.000  & I301\_76K\_CHARMS  &     125.00  &           0 & Filter\\
  24 & STANDARD  &    Infinity  &      10.000  &                    &          -  &           0 & \\
  25 & STANDARD  &    Infinity  &     163.500  &                    &     100.00  &           0 & Lyot Stop\\
  26 & STANDARD  &     304.883  &      37.478  & CAF2\_76K\_CHARMS  &     163.59  &           0 & Lens 1 Camera\\
  27 & STANDARD  &    -325.003  &      20.653  &                    &     163.59  &           0 & \\
  28 & STANDARD  &    -253.737  &       6.979  &  S-FTM16\_L10\_76  &     159.85  &           0 & Lens 2 Camera\\
  29 & STANDARD  &    Infinity  &      10.739  &                    &     159.85  &           0 & \\
  30 & STANDARD  &     262.371  &      34.005  & CAF2\_76K\_CHARMS  &     162.18  &           0 & Lens 3 Camera\\
  31 & STANDARD  &    Infinity  &     187.848  &                    &     162.18  &           0 & \\
  32 & STANDARD  &     192.373  &      21.201  & BAF2\_76K\_CHARMS  &     137.44  &           0 & Lens 4 Camera\\
  33 & STANDARD  &    -966.735  &      29.159  &                    &     137.44  &           0 & \\
  34 & STANDARD  &     160.885  &      38.080  & CAF2\_76K\_CHARMS  &     116.85  &           0 & Lens 5 Camera\\
  35 & STANDARD  &     757.961  &      23.302  &                    &     116.85  &           0 & \\
  36 & STANDARD  &    -144.813  &       8.071  &  S-FTM16\_L14\_76  &      84.47  &           0 & Lens 6 Camera\\
  37 & STANDARD  &     154.348  &      23.172  &                    &      84.47  &           0 & \\
 IMA & STANDARD  &    Infinity  &              &                    &      52.13  &           0 & Detector\\
\enddata
\tablenotetext{a} {Aspheric coefficients: a$_4$=-5.2750111$\times$10$^{-8}$, a$_6$=2.1900236$\times$10$^{-12}$,
a$_{8}$=-7.7923832$\times$10$^{-17}$}
\end{deluxetable}

\begin{deluxetable}{rrrr}
\tabletypesize{\scriptsize}
\tablecaption{Filters\label{table:filters}}
\tablewidth{0pt}
\tablehead{
\colhead{Band} & \colhead{50\% Transmission Limits} & \colhead{Peak Transmission}
& \colhead{Vendor}
}
\startdata
zJ        & 0.949-1.500 &  0.94 & Materion-Barr\\
Y         & 0.966-1.072 &  0.94 & Research Electro-Optics\\
J         & 1.172-1.330 &  0.92 & Research Electro-Optics\\
H         & 1.498-1.786 &  0.96 & Research Electro-Optics\\
HK        & 1.253-2.489 &  0.92 & Omega Optical\\
K$_s$     & 1.996-2.318 &  0.86 & Research Electro-Optics\\
K$_{spec}$& 1.931-2.451 &  0.98 & Omega Optical\\
\enddata
\end{deluxetable}

\begin{deluxetable}{rrrrr}
\tabletypesize{\scriptsize}
\tablecaption{Dispersers\label{table:dispersers}}
\tablewidth{0pt}
\tablehead{
\colhead{Band} & \colhead{lines mm$^{-1}$} & \colhead{Prism Angle In} & \colhead{Prism Angle Out}
& \colhead{Prism Material}
}
\startdata
J (ruled)  & 210 &   0$^{\circ}$ & 20.1$^{\circ}$ & Infrasil\\
H (ruled)  & 150 & 9.2$^{\circ}$ & 26.7$^{\circ}$ & Infrasil\\
HK (ruled) & 81.6& 9.2$^{\circ}$ & 26.7$^{\circ}$ & Infrasil\\
H (VPH)    & 290 & 9.5$^{\circ}$ &  9.5$^{\circ}$ & ZnSe$^a$\\
K (VPH)    & 200 & 8.8$^{\circ}$ &  8.8$^{\circ}$ & ZnSe$^a$\\
\enddata
\tablenotetext{a} {VPH gratings use a pair of identical prisms.}
\end{deluxetable}

\begin{deluxetable}{rrrrrr}
\tabletypesize{\scriptsize}
\tablecaption{Spectroscopic Modes\label{table:modes}}
\tablewidth{0pt}
\tablehead{
\colhead{Wavelength Range} & \colhead{Disperser} & \colhead{Order} & \colhead{Filter}
& \colhead{Pixels/Spectrum} & \colhead{Resolution}
}
\startdata
0.96-1.07 & HK &      2 &   Y & 360& 1600 	\\
0.96-1.07 & J  &      1 &   Y & 520& 2400 	\\
0.96-1.07 & H  &      2 &   Y & 670& 3000 	\\
0.95-1.50 & HK &    1+2 &  zJ & var&  800,1600 \\
0.94-1.51 & J  &      1 &  zJ &2600& 2400 	\\
1.17-1.33 & J  &      1 &   J & 720& 2800 	\\
1.25-2.45 & HK &      1 &  HK &1800& 1400 	\\
1.50-1.79 & H  &      1 &   H & 800& 2400 	\\
1.50-1.79 & H (VPH) & 1 &	H &    & 3000 \\
1.98-2.32 & HK &      1 &   K & 550& 1700 	\\
1.95-2.45 & K (VPH) & 1 &Kspec&    & 3000 \\
\enddata
\end{deluxetable}

\begin{deluxetable}{llll}
\tabletypesize{\scriptsize}
\tablecaption{Mechanisms\label{table:mechanisms}}
\tablewidth{0pt}
\tablehead{
\colhead{Mechanism} & \colhead{Mechanism Type} & \colhead{Gearing} & \colhead{Position Sensing}
}
\startdata
Calibration Mirror & ballscrew linear  & planetary    & Hall effect limits/home\\
Guiders         & 3 ballscrew linear   & planetary    & Hall effect limits/home\\
                & (positioning)        &              & magnetic tape encoder \\
                & 1 cam driven linear  & planetary    & magnetic tape encoder\\
                & (focus)              &              &     \\
Dekker Wheel    & rotary with sapphire & worm gear at & microswitch detent detection/home\\
                & thrust bearing       & outer radius &     \\
Slit Mask Wheel & as Dekker            & worm gear at & as Dekker\\
                &                      & 40\% radius  &  \\
Gate Valve      & ballscrew linear     & 90$^{\circ}$ worm gear & microswitch/reed switch limits/home\\

Disperser Wheel & as Dekker            & as Dekker    & as Dekker\\
Filter Wheels   & as Dekker            &              & as Dekker\\
Focus Stage     & cam driven linear    & pinion/spur  & LVDT\\
\enddata
\end{deluxetable}

\begin{deluxetable}{ll}
\tabletypesize{\scriptsize}
\tablecaption{HAWAII 2 operating characteristics. \label{table:array}}
\tablewidth{0pt}
\tablehead{ \colhead{Property} & \colhead{Typical Performance} }
\startdata
Gain                           &  5 $e^-$ DN$^{-1}$ \\
Read noise (double correlated read)       & 17 e-    \\
Read noise (60 reads)          &  5 $e^-$ \\
Full well                      & 	230,000 e- or 46,000 DN \\
Linearity correction           & 	usable to 41,000 DN (90\% of full well) \\
Dark current                   & 	0.1 to 0.3 $e^-$ s$^{-1}$ pix$^{-1}$ at 78 K   \\
Readout time  	               & 32 channels at 180 kHz: 0.7 sec per read \\
Multiple sampling mode         & "up-the-ramp" \\
\enddata
\end{deluxetable}

\begin{deluxetable}{lllll}
\tabletypesize{\scriptsize}
\tablecaption{Imaging characteristics \label{table:imageprop}}
\tablewidth{0pt}
\tablehead{ \colhead{Property} & \colhead{Y} & \colhead{J} & \colhead{H} & \colhead{K} }
\startdata
System Throughput             & 0.25 & 0.22 & 0.29 & 0.26 \\
Time to reach 50\% saturation &  600 &  400 &  50 &    65 \\
\enddata
\end{deluxetable}

\begin{deluxetable}{rllll}
\tabletypesize{\scriptsize}
\tablecaption{Imaging Vega magnitude limit for signal/noise of 10. \label{table:imagesens}}
\tablewidth{0pt}
\tablehead{ \colhead{Integration Time in s} & \colhead{Y} & \colhead{J} & \colhead{H} & \colhead{K} }
\startdata
60 	    & 20.3 & 20.1 & 19.5 & 18.8 \\
300     & 21.2 & 21.0 & 20.4 & 19.7 \\
3600    & 22.5 & 22.4 & 21.7 & 21.0 \\
\enddata
\end{deluxetable}

\begin{deluxetable}{rllll}
\tabletypesize{\scriptsize}
\tablecaption{Spectroscopic Vega magnitude limit for signal/noise of 10 per resolution element at R=3000 between OH sky lines. \label{table:specsens}}
\tablewidth{0pt}
\tablehead{ \colhead{Integration Time in s} & \colhead{Y} & \colhead{J} & \colhead{H} & \colhead{K} }
\startdata
3600    & 20.2 & 20.2 & 20.3 & 18.4 \\
\enddata
\end{deluxetable}

\end{document}